\documentclass[a4paper,10pt]{IEEEtran}

\usepackage[T1]{fontenc}
\usepackage[utf8]{inputenc} 
\usepackage{amsmath,amssymb}
\usepackage{graphicx}
\usepackage{float}
\usepackage{caption}
\usepackage{subcaption}
\usepackage{booktabs}
\usepackage{multirow}
\usepackage{tabularx}
\usepackage{array}

\usepackage{algorithm}
\usepackage{algpseudocode}

\usepackage[table]{xcolor}  
\usepackage{colortbl}

\usepackage{tikz}

\usepackage{hyperref}

\usepackage{comment}
\usepackage{authblk}

\definecolor{lime}{HTML}{A6CE39}
\DeclareRobustCommand{\orcidicon}{%
    \begin{tikzpicture}
    \draw[lime, fill=lime] (0,0) 
    circle [radius=0.16] 
    node[white] {{\fontfamily{qag}\selectfont \tiny ID}};
    \draw[white, fill=white] (-0.0625,0.095) 
    circle [radius=0.007];
    \end{tikzpicture}
    \hspace{-2mm}
}

\newcommand{\orcidauthorA}{0000-0003-3148-3765} 
\newcommand{\orcidauthorB}{0009-0006-8733-4119} 
\newcommand{\orcidauthorC}{0000-0003-3884-9520}

\newcommand{\orcid}[1]{\href{https://orcid.org/#1}{\orcidicon}}

\title{ForgeMorph: An FPGA Compiler for On-the-Fly Adaptive CNN Reconfiguration}

\author[1]{Mazouz Alaa Eddine\orcid{\orcidauthorA}}
\author[1]{Duc Han Le\orcid{\orcidauthorB}}
\author[1]{Van Tam Nguyen\orcid{\orcidauthorC}}

\affil[1]{LTCI, Télécom Paris, Institut Polytechnique de Paris, France}

\affil[]{Email: alaa.mazouz@telecom-paris.fr}

\date{March 2025}

\begin{document}
\maketitle

\begin{abstract}
We introduce \textbf{ForgeMorph}, a full-stack compiler for adaptive CNN deployment on FPGAs, combining design-time optimization with runtime reconfigurability. At compile time, the \textit{NeuroForge} engine performs constraint-driven design space exploration, generating RTL mappings that are Pareto-optimal with respect to user-defined latency and resource budgets. Unlike existing FPGA compilers, which rely on static scheduling and manual tuning, NeuroForge leverages analytical performance models and multi-objective genetic algorithms to efficiently search large configuration spaces and propose highly optimized hardware implementations.
At runtime, the \textit{NeuroMorph} module enables dynamic reconfiguration of network width and depth without requiring redeployment. This is made possible by a novel training strategy, \textit{DistillCycle}, which jointly trains the full model and its subnetworks using hierarchical knowledge distillation. As a result, each execution path maintains accuracy even under aggressive resource and power constraints.
We demonstrate ForgeMorph on the Zynq-7100 using custom and benchmark models including MobileNetV2, ResNet-50, SqueezeNet, and YOLOv5. The system achieves up to \textbf{50$\times$ latency reduction} and \textbf{32\% lower power consumption} at runtime, while matching or exceeding the efficiency of state-of-the-art compilers. ForgeMorph offers a unified solution for deployment scenarios that demand flexibility, performance, and hardware efficiency.
\end{abstract}

\section{Introduction} 
\label{introduction} 
Emerging AI systems—ranging from autonomous agents like self-driving cars to real-time embedded applications—require rapid CNN inference under strict latency and power constraints. These systems operate in time-sensitive environments where DNN outputs must inform actions with limited compute. For example, an autonomous vehicle performing object detection must deliver predictions fast enough to guide real-time control. Similarly, mobile devices may enter power-saving modes that require reduced computation without compromising functionality. Spaceborne platforms such as satellites and planetary rovers~\cite{stimpson2017,mazouz2019,AEM2024} also benefit from reconfigurable, reliable DNN implementations~\cite{kouris2019,Chai2023}, as demonstrated in tasks like visual feature detection~\cite{santos2020} and traffic sign recognition~\cite{Habeeb2023,yih2018}.

While DNNs excel in accuracy, their high computational and energy demands pose deployment challenges on embedded platforms with tight performance and power budgets~\cite{Han2024}. A key limitation in the current hardware deep learning landscape is the lack of support for online reconfiguration and adaptivity. Enabling runtime flexibility would broaden the scope of applications and allow systems to trade off accuracy, latency, and energy dynamically.

Adaptive deep models~\cite{Yaping2019}, combined with FPGA-based partial reconfiguration, offer a promising solution—allowing hardware to adjust its execution profile without disrupting critical operation. Yet this capability remains largely unused, as existing frameworks lack support for generic CNN modules and reconfiguration-ready architectures capable of scalable, adaptive inference. Realizing this potential requires compiler-level automation to generate optimized hardware implementations from high-level specifications. Leveraging parameterized, pre-built libraries tailored to user-defined constraints would significantly accelerate hardware design, improve portability, and remove the need for manual RTL development.

To address these limitations, we present \textbf{ForgeMorph}—a unified compiler framework that integrates automated design-time exploration with dynamic runtime reconfiguration. ForgeMorph targets application-specific constraints on latency, energy, and accuracy through a two-stage optimization strategy.

The framework comprises two complementary components: \textbf{NeuroForge}, a design space exploration engine that generates optimized RTL from high-level CNN specifications, and \textbf{NeuroMorph}, a runtime module that enables on-the-fly adaptation of inference pipelines. NeuroForge uses a multi-objective genetic algorithm to explore loop unrolling, pipelining, and PE allocation, producing stream-based, fully pipelined architectures suited for real-time operation. The exploration begins with coarse Simulink-based latency and area estimates and refines candidate configurations through RTL synthesis and block-level profiling.

At runtime, NeuroMorph enables selective activation of embedded subnetworks via clock gating, allowing the system to dynamically scale its compute footprint without resynthesis or retraining. This is made possible by an alternating training strategy that ensures all subnetworks are jointly optimized for functional correctness and performance. Transitions between configurations are controlled via lightweight toggles, supporting real-time trade-offs across accuracy, power, and latency modes.

\begin{figure*}[htbp]
  \centering
  \includegraphics[width=0.6\textwidth]{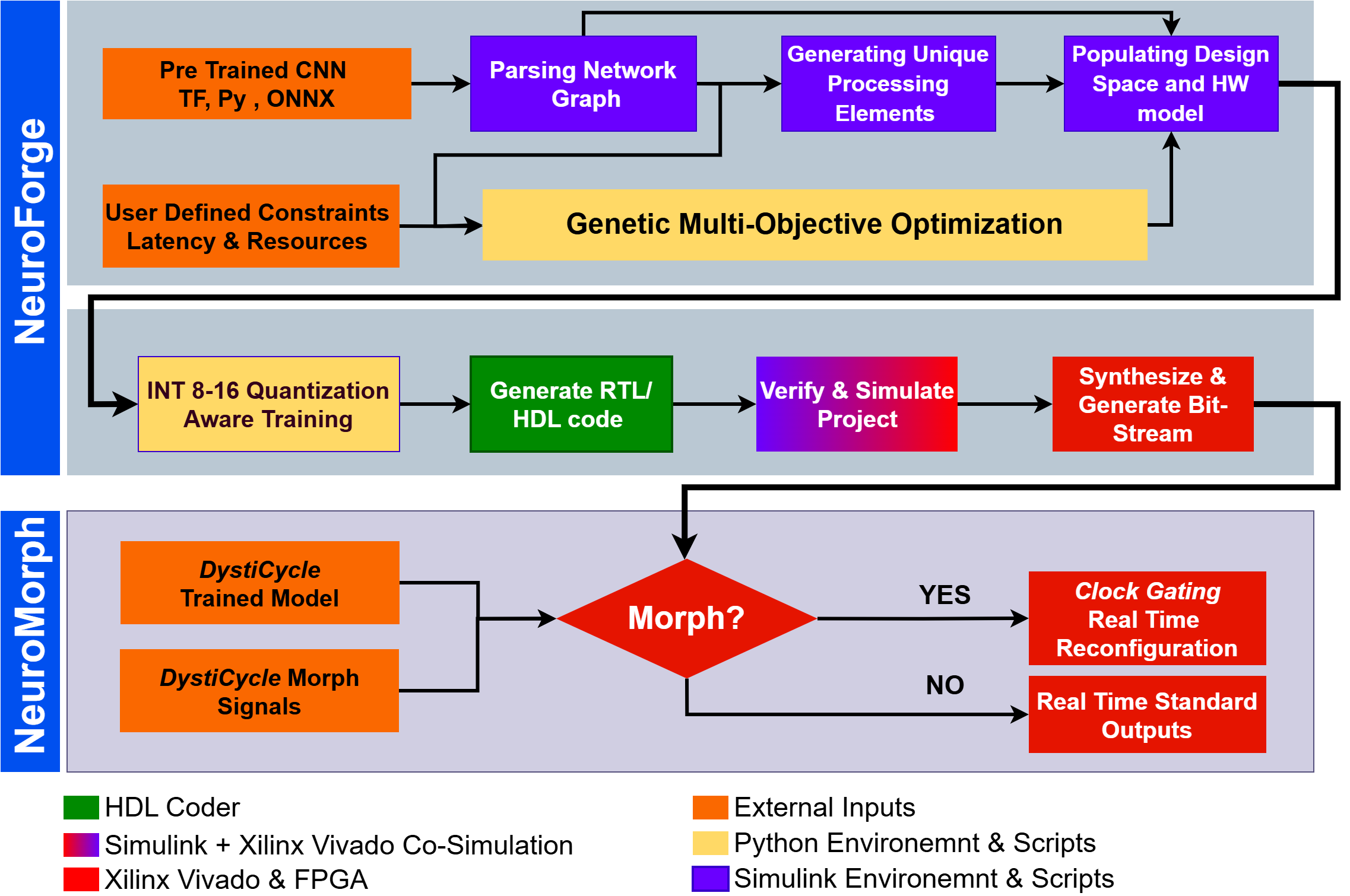}  
    \caption{Diagram of the proposed \textbf{NeuroForge} and \textbf{NeuroMorph} system, illustrating the full \textbf{ForgeMorph} compiler workflow for mapping, compiling and adaptively deploying models on FPGA.}
    \label{fig:worflow}
\end{figure*}

The full ForgeMorph pipeline is illustrated in Fig.~\ref{fig:worflow}, highlighting the transformation from high-level model descriptions to adaptive hardware-ready implementations.

\begin{figure}[b]
  \centering
  \includegraphics[width=0.8\columnwidth]{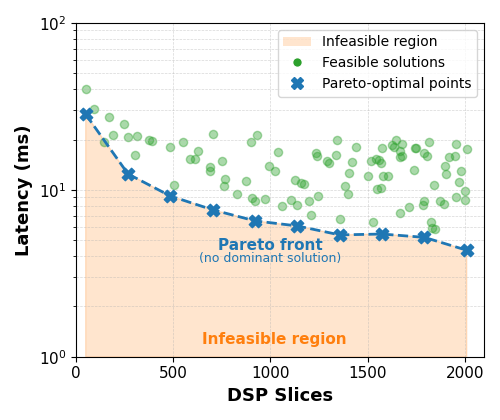}
  \caption{
 Illustration of \textbf{NeuroForge}'s design space exploration for a CIFAR-10 model, showing the Pareto front of latency versus resource usage across candidate hardware mappings.  
It enables rapid, customizable exploration by generating optimized configurations that satisfy user-defined performance and resource constraints.  
Each point represents a valid design, with the Pareto front highlighting optimal trade-offs for guiding design-time decisions.
  }
  \label{fig:ODE_results}
\end{figure}

\begin{figure}[htbp]
  \centering
  \includegraphics[width=1\columnwidth]{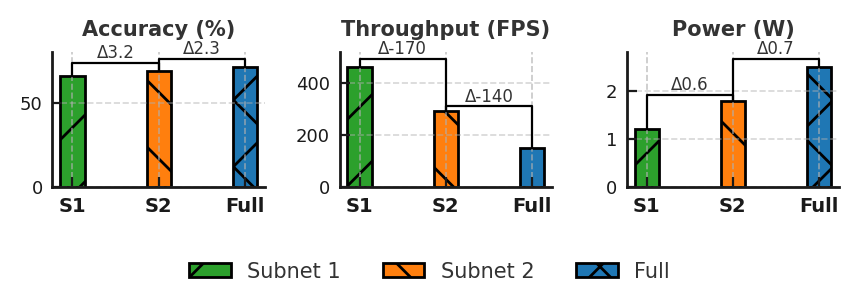}
  \caption{
Illustration of \textbf{NeuroMorph}'s runtime reconfigurability using a VGG16  model trained on ImageNet with \textbf{DystillCycle}. The network is partitioned into two subnets and the full model, enabling adaptive trade-offs between accuracy, throughput, and power. At runtime, the system can switch configurations to meet performance or energy constraints.
  }
  \label{fig:Morph_example}
\end{figure}

\textbf{Our key contributions are:}
\begin{itemize}
    \item \textbf{A fully automated, Simulink-driven design stack for RTL generation}, capable of rapidly compiling convolutional neural networks into high-performance FPGA implementations. The framework integrates hardware-aware modeling and parameter inference to support platform-specific optimization without manual tuning.

    \item \textbf{A fast and analytically driven design space exploration engine, \textit{NeuroForge}}, which couples coarse-grained analytical estimations with a multi-objective genetic algorithm to optimize unrolling, tiling, and resource allocation, example seen in Fig.~\ref{fig:ODE_results}. NeuroForge enables accurate and scalable exploration without requiring full RTL simulation or time-consuming synthesis in the loop.

    \item \textbf{A runtime reconfiguration module, \textit{NeuroMorph}}, that supports both width and depth-wise decomposition of models into selectively activated subnetworks, controlled via clock gating. Unlike prior approaches, NeuroMorph does not require duplicating entire models or reprogramming the FPGA, offering true low-overhead, on-chip adaptability, an example can be seen in Fig.~\ref{fig:Morph_example}.

    \item \textbf{A novel training methodology, \textit{DistillCycle}}, which uses cyclic distillation and alternating subnetwork activation to ensure that all reconfigurable paths (including early exits) maintain functional accuracy. This allows models to degrade gracefully and reliably under runtime constraints, enabling smooth trade-offs between accuracy, latency, and energy.

    \item \textbf{The first integrated FPGA compiler framework, \textit{ForgeMorph}, to combine design-time exploration with runtime adaptivity}, offering both fine-grained control over hardware mappings and post-deployment flexibility. ForgeMorph enables developers to generate reconfigurable hardware designs directly from high-level CNN models with minimal manual effort.
\end{itemize}

 The remainder of this paper is organized as follows: Section~\ref{NeuroForge} presents \textit{NeuroForge}, our offline design space exploration engine. Section~\ref{NeuroMorph} introduces \textit{NeuroMorph}, our runtime reconfiguration framework. Section~\ref{expermental} reports experimental results and comparisons. Section~\ref{sec:conclusions} concludes and outlines future work.

\section{Related Work} \label{related_work} FPGAs are well-suited for real-time, compute-intensive deep learning applications due to their power efficiency, high parallelism, and design flexibility~\cite{venkatesh2017}. Their modular hardware architecture enables rapid prototyping and allows selective deployment of design components at runtime without compromising overall system functionality—an essential feature given the rapid evolution of CNN architectures.

Earlier work primarily focused on implementing simple neural networks in hardware, often neglecting scalability to more complex models~\cite{venkatesh2017,nurvitadhi2017}. Recent efforts have shifted toward optimizing (1) model parameters and architecture at the software level, and (2) storage and memory organization at the hardware level, particularly for embedded systems. Additionally, some studies have explored automating the CNN-to-FPGA deployment process~\cite{gan2016}.

\subsection{Design Space Exploration (DSE)}

Many recent frameworks aim to automate the mapping of DNNs onto FPGA platforms by exploring architectural design choices such as loop unrolling, tiling, and resource reuse. However, most of these frameworks rely on \textit{slow}, \textit{simulation-heavy}, or \textit{heuristic-based exploration}, and offer limited visibility into the design process.

Tools like \textbf{DNNBuilder}~\cite{DNNBuilder}, \textbf{AutoDNNchip}~\cite{AutoDNN}, \textbf{DNNExplorer}~\cite{DNNExplorer}, and \textbf{DeepBurning}~\cite{DeepBurning} build accelerators using parameterized templates or high-level synthesis (HLS). These systems support unrolling, pipelining, and operator customization, but often depend on exhaustive architecture simulation or full RTL synthesis for performance estimation. As a result, they are computationally expensive, slow to iterate, and difficult to tune—especially for time-sensitive or embedded applications.
For example, \textbf{DNNBuilder} introduces a pipelined architecture and uses a resource-guided scheduling strategy, but its performance estimation depends on empirical profiling and hand-crafted templates. \textbf{AutoDNNchip} and \textbf{DNNExplorer} provide efficient chip-level search spaces and performance predictors, but rely on black-box models and multi-stage build flows that may be slow or inflexible. \textbf{DeepBurning} generates RTL from neural models but lacks dynamic reconfiguration and cannot easily target hardware-specific primitives such as embedded DSP chains.

Other solutions like \textbf{DNNWeaver} and \textbf{FINN-R}~\cite{FINN-R} emphasize interpretability and synthesis throughput, but are constrained by fixed IP templates or quantization assumptions. While these frameworks offer acceptable performance-resource trade-offs, they provide \textit{limited flexibility}, \textit{no support for runtime adaptivity}, and \textit{restricted control over mapping decisions}.

In contrast, our proposed \textbf{NeuroForge} design space exploration engine within the \textbf{ForgeMorph} framework offers a \textbf{fast}, \textbf{customizable}, and \textbf{fully automated} approach to DSE. It achieves this by:
\begin{itemize}
    \item Leveraging \textit{analytical models}, derived from high-level Simulink-based estimations, to evaluate latency and area trade-offs without full RTL synthesis.
    \item Applying a \textit{multi-objective genetic algorithm} to jointly optimize latency, energy, and resource utilization while exploring scheduling strategies such as loop unrolling and PE allocation.
    \item Supporting a \textit{transparent and modular design flow}, giving users control over performance targets while maintaining portability across FPGA devices.
\end{itemize}

Moreover, unlike existing DSE tools, ForgeMorph uniquely supports runtime adaptivity via the \textbf{NeuroMorph} module, enabling on-device clock-gated subnetwork reconfiguration. This allows dynamic transitions between performance modes without retraining or resynthesis—an ability not found in any prior compiler framework.
In summary, while prior DSE methods have contributed foundational tools for FPGA accelerator generation, they fall short in speed, transparency, and flexibility. ForgeMorph bridges this gap by offering fast analytical exploration paired with runtime reconfiguration, empowering users to deploy CNNs that are both performance-aware and responsive to real-world deployment constraints.

\subsection{Runtime Adaptive Reconfiguration} \label{sec:runtime_reconfiguration}

Runtime reconfiguration remains an underexplored dimension in FPGA-based deep learning, despite its critical role in enabling adaptive behavior under changing execution conditions. Inference workloads frequently operate under fluctuating power budgets, latency constraints, or dynamic accuracy requirements. Supporting flexible execution modes at runtime can significantly improve system responsiveness, power efficiency, and robustness in edge and embedded contexts.

Only a handful of works attempt to address this challenge. One of the earliest efforts, \textbf{CascadeCNN} \cite{kouris2018}, adopts a dual-model system that switches between low- and high-precision paths depending on confidence thresholds. Inspired by the “Big/Little” design pattern~cite{park2015}, it enables coarse-grained accuracy–throughput trade-offs. However, it requires both networks to be present simultaneously on-chip, incurring high resource and power overhead.
Another line of work, \textbf{FpgaConvNet}~\cite{venieris2016fccm,venieris2018tnnls}, implements dynamic reconfiguration by partially reprogramming the FPGA between layers. While this improves utilization efficiency, it is impractical for runtime switching due to the long reconfiguration time and lack of selective granularity. These systems remain fundamentally static in their inference flow, unable to react fluidly to real-time performance constraints.
More recently, ~\cite{predictive_exit} introduced early-exit-based dynamic computation by integrating intermediate classifiers into the model pipeline. While it introduces runtime flexibility at the algorithmic level, it remains a design-time strategy with no FPGA-level runtime control, and does not account for the performance loss induced by early exiting during training. Its static early-exit points are fixed into the model without any training regularization to balance their outputs.

By contrast, \textbf{NeuroMorph}—our runtime component within the \textbf{ForgeMorph} compiler—offers true runtime reconfiguration in hardware. It supports depth-wise decomposition of networks into selectively activatable subnetworks, enabling on-the-fly performance scaling using a single jointly trained model. Reconfiguration is performed through clock-gated switching controlled by lightweight runtime logic, avoiding any need for full reprogramming or model duplication.

Critically, NeuroMorph’s flexibility is made possible by our custom training strategy, \textbf{DistillCycle}, which uses staged distillation and cyclic subnetwork activation to regularize both main and early-exit outputs. This ensures that all reconfigurable branches perform reliably and minimizes accuracy degradation compared to fixed-path networks. Unlike prior early-exit frameworks, our method couples training and hardware execution, maintaining both modularity and performance.

 Mainstream compilers such as \textbf{Vitis-AI} \cite{vitis_ai}, \textbf{OpenVINO} \cite{openvino}, \textbf{TVM} \cite{tvm}, and \textbf{hls4ml} \cite{hls4ml} offer powerful deployment pipelines but are confined to static inference workflows. They lack mechanisms for dynamically adjusting execution paths, balancing energy and accuracy, or selectively enabling compute regions based on runtime feedback. In contrast, ForgeMorph enables runtime adaptivity, modular execution, and design-aware training, offering a uniquely flexible alternative to existing compiler frameworks.

\textbf{NeuroMorph} is the first framework, to our knowledge, to integrate reconfigurability at the hardware level with a novel training methodology that combines \textit{knowledge distillation} and \textit{predictive early-exit clock gating}. This enables each subnetwork to be independently accurate, compact, and triggerable during inference—laying the foundation for truly adaptive, resource-aware DNN deployment.

\section{NeuroForge: Offline Design Exploration}
\label{NeuroForge}
\textit{NeuroForge} is the design-time exploration engine of the ForgeShift compiler, responsible for automatically generating platform-agnostic RTL code tailored to user-defined constraints such as latency, DSP usage, or throughput. At its core, NeuroForge leverages a Multi-Objective Genetic Algorithm (MOGA) to explore a wide range of hardware configurations—evaluating combinations of loop unrolling, pipelining, and PE replication—to identify Pareto-optimal trade-offs between performance and resource utilization.
The Model-Based Design (MBD) workflow within NeuroForge begins by parsing a pre-trained network graph to extract topology, layer types, and parameter information. It supports both sequential and residucal CNNs. Based on the parsed architecture, the system generates specialized Processing Elements (PEs) for core operations. These PEs are then composed into candidate hardware models whose performance and area are estimated using analytical models. Once validated functionally, the most promising designs are selected for RTL synthesis and deployment.
The following sections detail each stage of this workflow as seen in Fig.~\ref{fig:worflow}.

\subsection{Parsing Network Graphs and Generating Processing Elements}

NeuroForge’s design exploration begins by parsing pre-trained network graphs from formats such as MATLAB, TensorFlow, PyTorch, and ONNX. The parser extracts layer topology and parameters—such as filter count ($N$), kernel size ($K$), stride ($S$), padding ($P$), and input dimensions (\( FM_i^{H} \), \( FM_i^{W} \), \( Ch^{D} \) for convolutions, or input/output size ($\text{FC}_{\text{In}}$, $\text{FC}_{\text{Out}}$) for fully connected layers. Residual blocks are interpreted as subgraphs of convolutional layers with skip connections; their main and shortcut paths are fused into modular blocks based on graph connectivity.

Connection topologies are captured via a connection table specifying source-to-destination layer mappings. While sequential CNNs follow a strict chain, residual architectures feature skip connections and multi-branch paths. The parser identifies convergence points—where skip and main branches merge—and represents these as \textit{ResidualAdd} layers, later synthesized into arithmetic units.

Once the network is parsed, NeuroForge generates generic Processing Elements (PEs) and builds Processing Units (PUs) for each functional layer using a library of hardware models in Simulink. Layer parameters such as kernel size and stride guide the configuration of each PE.

\subsubsection{Convolutional PEs \( (C_{\text{PE}} \))}

Each convolutional layer is implemented as a set of \(\text{C}_{\text{PE}}\) units, each comprising a two-stage pipeline: 

\begin{itemize}
    \item \textbf{Line Buffer Controller (LBC)} – Buffers input rows and assembles local windows for convolution.
    \item \textbf{Multiply-Accumulate (MAC) Core} – Executes element-wise multiplication and accumulation.
\end{itemize}

\begin{figure}[htbp]
    \centering
    \includegraphics[width=0.65\columnwidth]{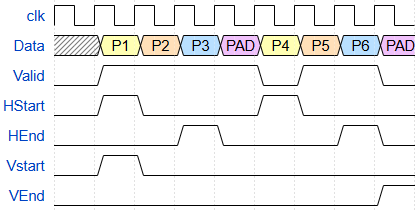}
    \caption{Control signal encoding for a $3\times2$ window with one-pixel padding \cite{mazouz2020, mazouz2019ahs}.}
    \label{fig:control_signal}
\end{figure}

To support streaming data, each input pixel is accompanied by a 5-bit control signal (\texttt{Valid}, \texttt{hStart}, \texttt{hEnd}, \texttt{vStart}, \texttt{vEnd}) that encodes spatial context for window generation, as illustrated in Fig.~\ref{fig:control_signal}.
The Line Buffer Controller consists of a series of FIFOs that shift data according to stride \( S \). To form valid \( n \times m \) convolution windows, it buffers \( n{-}1 \) full rows and collects \( n \) elements from each of \( m \) FIFOs into a register bank.

\begin{figure}[htbp]
  \centering
  \includegraphics[width=0.60\columnwidth]{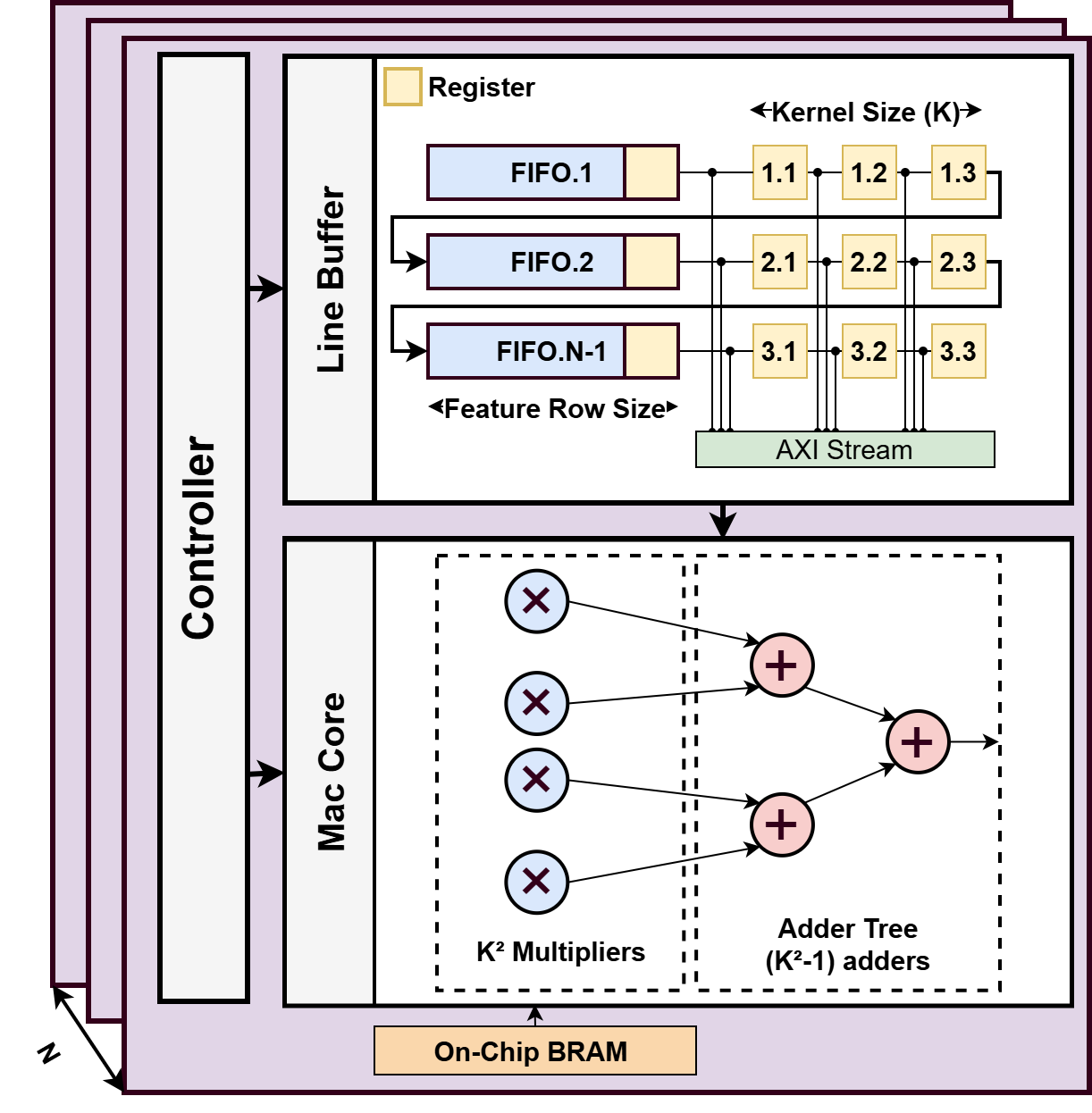}
  \caption{Convolutional PE array with $N$ parallel units, each using a $3\times3$ kernel.}
  \label{fig:conv_pe}
\end{figure}
Each valid window is passed to the MAC core, which performs \( K^2 \) parallel multiplications followed by a multi-level adder tree to accumulate the result. The number of multipliers and the depth of the adder tree are given by:
\begin{equation}
N_{\mathrm{mult}} = K^2,
\label{eq:num_mult}
\end{equation}
\begin{equation}
\max(N_{\mathrm{add\_stages}}) = \left\lceil \log_2(K^2) \right\rceil + 1,
\label{eq:max_adder_stages}
\end{equation}
\begin{equation}
N_{\mathrm{add}} = \frac{(K^2 - 1) \times N_{\mathrm{add\_stages}}}{\max(N_{\mathrm{add\_stages}})}.
\label{eq:num_adders}
\end{equation}

For example, a \( 3 \times 3 \) kernel results in 9 multipliers and 8 adders across 5 pipeline stages, enabling one output per clock cycle.

Each \( C_{\text{PE}} \) MAC core includes \((FM_i^{H} - 1)\) FIFOs of size \( FM_i^{W} \), \( K^2 \) multipliers, \( K^2 \) registers, and \( K^2 - 1 \) adders. The memory control unit uses \( K \) adders for address generation. A ReLU stage applies a comparator-based non-linearity, introducing one clock cycle per element. Fig.~\ref{fig:conv_pe} illustrates a convolutional unit with \( N \) parallel \( C_{\text{PE}} \) units for \( K=3 \).
We denote by \( \tau_{C_{\text{PE}}} \) the latency of a single processing element, which consists of the core pipeline delay and any additional overhead from control logic.

\begin{equation}
\tau_{C_{\text{PE}}} = \text{Clk} \times L_{\text{core}} + T_{\text{overhead}},
\label{eq:cpe_latency}
\end{equation}

where
\begin{align*}
L_{\text{core}} &= D_{\text{in}} + \frac{P_b + 1}{2} + (W + P_b + P_f) \times H, \\
T_{\text{overhead}} &= T_{\text{pad}} + T_{\text{tap}} + T_{\text{mul}} + T_{\text{add}} + D_{\text{out}} + T_{\text{ReLU}}.
\end{align*}

\begin{itemize}
    \item \( W, H \): Input feature map width and height.
    \item \( P_b, P_f \): Back and front porch (blanking intervals).
    \item \( D_{\text{in}}, D_{\text{out}} \): Input/output delays (4 cycles each; \( D_{\text{in}} \) only for the first layer).
    \item \( T_{\text{pad}} \): Padding latency.
    \item \( T_{\text{tap}} \), \( T_{\text{mul}} \): Delays for tapping the line buffer and loading multipliers (each \( K \) cycles).
    \item \( T_{\text{add}} \): Adder tree latency (\( N_{\text{Clk}} + 2 \) cycles).
    \item \( T_{\text{ReLU}} \): Delay for ReLU activation (1 cycle per element).
\end{itemize}

\subsubsection{Pooling PEs}

Average pooling reuses the \( C_{\text{PE}} \) structure but replaces learned weights with fixed coefficients, eliminating the need for weight registers or additional memory reads. Max pooling also shares the same memory controller but substitutes the MAC core with a \( K^2 \)-comparator tree to extract the maximum value within each window.

\subsubsection{Fully Connected PEs  (FC\(_\text{PE}\))}

The fully connected processing element (FC\(_\text{PE}\)) contains a MAC unit that computes the weighted sum of streamed inputs. Each input is multiplied by its corresponding weight and accumulated in an output register. Once all input-weight pairs are processed, the result is stored—repeating this for all \( FC_{\text{Out}} \) output heads.
Each head uses one multiplier and two adders, yielding \( FC_{\text{Out}} \) multipliers and \( 2 \times FC_{\text{Out}} \) adders for the layer. Unlike convolutional layers that exploit spatial parallelism, FC layers require vectorized input, introducing a serial processing bottleneck. This limits throughput as data is streamed element-wise. The computation is formalized in Eq.~(\ref{eq:fc_basic}).

\begin{equation}
Y(i) = \left[ Y_{i1} + Y_{i2} + \cdots + Y_{in} \right] + B_i.
\label{eq:fc_basic}
\end{equation}

\noindent Here, \( Y(i) \) is the \( i \)-th output activation, where each term \( Y_{ij} = W_{ij} \cdot X_j \) represents the product of the \( j \)-th input feature \( X_j \) and its corresponding weight \( W_{ij} \), summed across all \( n \) input features and offset by a bias term \( B_i \).

To alleviate the serialization bottleneck, we introduce channel-wise parallelism by avoiding full vectorization. Instead, each input channel is processed independently by parallel FC-Accumulation blocks. Final output values are obtained by aggregating partial sums from all input channels, as described in Eq.~(\ref{eq:fc_parallel}).

\begin{align}
Y(i) = \bigg( 
& \sum_{j=1}^{n} W_{i1} x_j + \cdots + \sum_{j=1}^{n} W_{in} x_j.
\bigg) + B_i.
\label{eq:fc_parallel}
\end{align}

\noindent The high-level resource requirements for the FC\(_{\text{PE}}\) are then given by:

\begin{equation}
N_{\mathrm{mult}} = FC_{\mathrm{out}} \times N,
\label{eq:fc_mult}
\end{equation}

\begin{equation}
N_{\mathrm{add}} = \left( FC_{\mathrm{out}} \times N \right) + \left( FC_{\mathrm{out}} \times L \right),
\label{eq:fc_add}
\end{equation}

\begin{equation}
N_{\mathrm{reg}} = FC_{\mathrm{out}} \times N,
\label{eq:fc_reg}
\end{equation}

\noindent where \( L \) is the number of adders in the adder tree, and \( N \) is the number of FC\(_\text{PE}\) units allocated per output head.

When weights and biases are preloaded, each MAC operation takes one clock cycle per input element. The final latency of the FC layer is determined by the time required to stream in all input channels. We denote the latency of the fully connected processing element as \( \tau_{\text{FC}_{\text{PE}}} \), defined in Eq.~(\ref{eq:fc_latency}).

\begin{equation}
\tau_{\text{FC}_{\text{PE}}} = \text{Clk} \times 
\left[
\begin{aligned}
& (FM_i^W + BP + FP) \\
& \quad \times \left( FM_i^H - 1 \right) + FM_i^H
\end{aligned}
\right]
\times P,
\label{eq:fc_latency}
\end{equation}

\noindent
where \( P \) is the parallelism coefficient, defined as the ratio of input channels to available FC\(_\text{PE}\) units:
\[
P = \frac{C_h^D}{FC_{\text{PE}}}.
\]

When \( P = 1 \), all channels are processed in parallel, minimizing latency at the cost of higher resource usage.
We exclude memory overhead from latency estimates to generalize the PE model, acknowledging a minor discrepancy between estimated and reported latencies. While insignificant for small designs, memory-read latency becomes more impactful in larger systems and should be included in future models.

\begin{figure}[htbp]
  \centering
  \includegraphics[width=0.8\columnwidth]{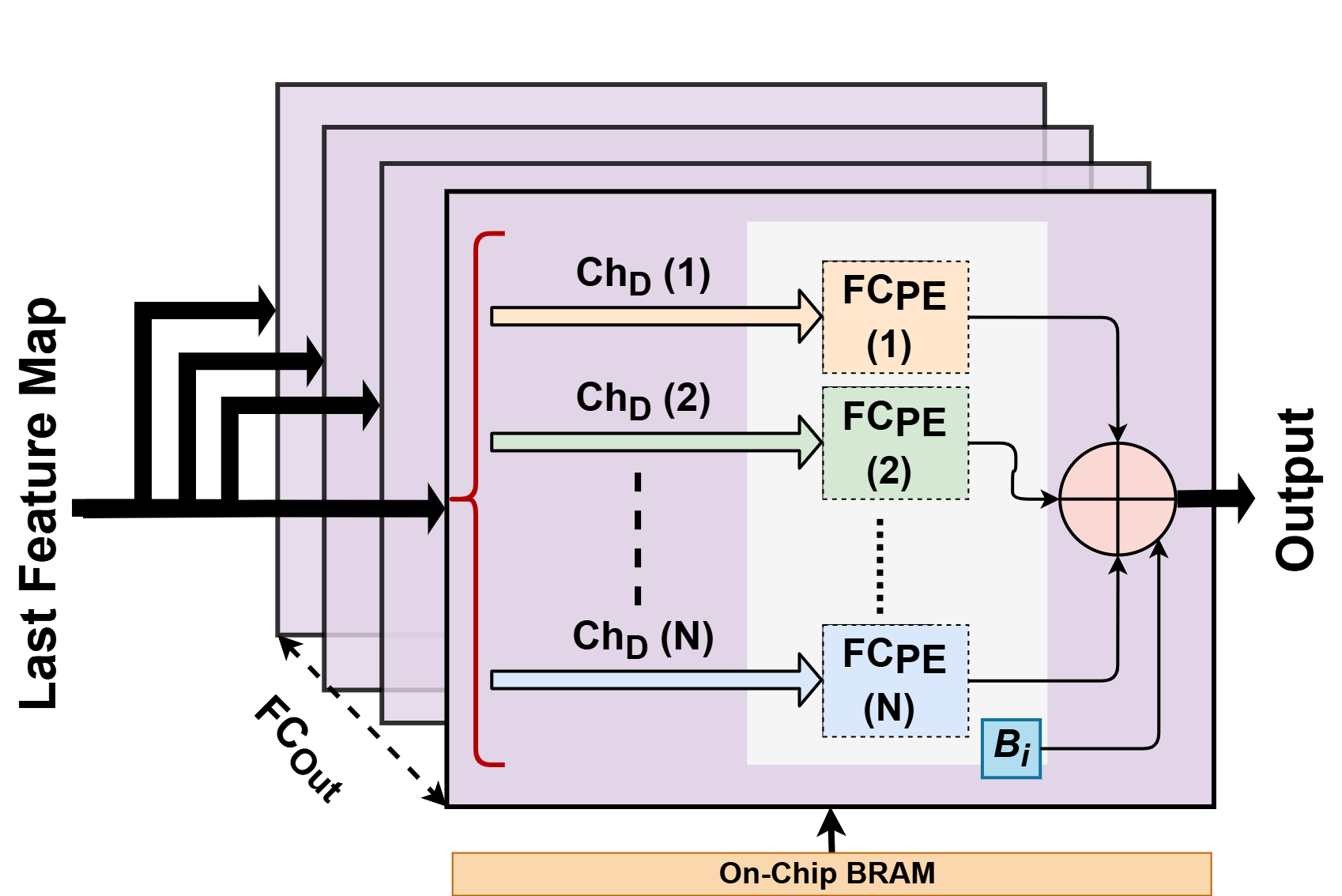}  
  \caption{4 Fully Connected Processing Unit with N Processing Elements.}
  \label{fig:FC_PE}
\end{figure}

\subsection{Populating Design Space and Hardware Model}

The hardware model representation follows the analytical and empirical estimations described previously, along with the resource models provided by Eq.~(\ref{eq:num_mult}) – (\ref{eq:fc_latency}).

\paragraph{\textbf{Convolutional Processing Elements \( C_{\text{PE}} \)}}
\begin{itemize}
    \item \emph{DSP slices:} Each \( C_{\text{PE}} \) maps \( K^2 \) DSP slices for MAC operations, where \( K \) is the kernel size. This ensures full parallelism across the convolution window.
    
    \item \emph{LUT slices:} Logic resources are primarily used for shift registers within the line buffer and for control logic. On average, each \( C_{\text{PE}} \) allocates around 800 LUTs, including approximately 300 dedicated to buffering.

    \item \emph{Block RAM (BRAM):} BRAM is used to store feature maps, weights, and intermediate results. 
    The required BRAM for the line buffer is estimated as:
\begin{equation}
    \text{BRAM}_{\text{linebuffer}} = \left\lceil \text{FM}_{\text{Size}} \times K \times \frac{\text{FP}_{\text{rep}}}{18\,\text{Kb}} \right\rceil,
\label{eq:bram_est}
\end{equation}

    where \( \text{FM}_{\text{Size}} \) is the feature map width, and \( \text{FP}_{\text{rep}} \) is the fixed-point representation bit width, NeuroForge supports both int8 and int16. 
\end{itemize}

\paragraph{\textbf{Pooling Processing Elements \( \text{PU}_{\text{PE}} \)}}
\begin{itemize}
    \item \emph{DSP slices:} Pooling operations do not require DSP slices, as they only involve comparisons or averaging, not multiplications.
    \item \emph{LUT slices:} Each \( \text{PU}_{\text{PE}} \) uses around 420 LUTs, primarily for logic and control, as well as for Line Buffer registers.
    \item \emph{Block RAM:} One BRAM per \( \text{PU}_{\text{PE}} \) is sufficient to store incoming elements and intermediate outputs.
\end{itemize}

\paragraph{\textbf{Fully Connected Processing Elements \( \text{FC}_{\text{PE}} \)}}
\begin{itemize}
    \item \emph{DSP slices:} Each \( \text{FC}_{\text{PE}} \) is assigned 1 DSP slice for MAC operations with input features.
    \item \emph{LUT slices:} Roughly 360 LUTs per \( \text{FC}_{\text{PE}} \) are used for logic and control.
    \item \emph{Block RAM:} \( \text{FC}_{\text{PE}} \) units do not require BRAM.
\end{itemize}

With all required \( C_{\text{PE}} \) components generated, NeuroForge constructs the hardware design space using the resource and latency models from Eq.~(\ref{eq:num_mult}) – (\ref{eq:bram_est}). These models guide the exploration of configurations by adjusting intra-layer parallelism and pipelining depth to trade off throughput and resource usage.
Each configuration maintains the original filter count but varies the number of active units and pipeline stages. Pooling, activation, and optional SoftMax layers are integrated inline for efficient streaming. The final selection is constrained by the target FPGA’s resource limits and operating frequency.
This stage forms the basis for the \textit{Design Space Exploration (DSE)} phase, introduced in the following section.

\subsection{DSE Using Multi-Objective Genetic Algorithm}

Design Space Exploration (DSE) is critical for mapping CNN workloads to FPGAs efficiently. The process is challenging due to the vast implementation space and the need to balance competing performance metrics under strict hardware constraints. Conventional approaches like Roofline Models (RLM)~\cite{siracusa2021} provide high-level performance estimates but do not generate concrete configurations tailored to user requirements or platform limits.

NeuroForge formulates DSE as a multi-objective optimization problem~\cite{cardoso2017,konak2006}, targeting two competing goals: reducing inference latency and minimizing resource utilization. These objectives often trade off against one another. The search space includes both intra-layer and inter-layer parallelism, and candidate configurations must satisfy constraints derived from the CNN architecture, the target FPGA platform, and any user-defined specifications.

\begin{figure}[htbp]
  \centering
  \includegraphics[width=0.7\columnwidth]{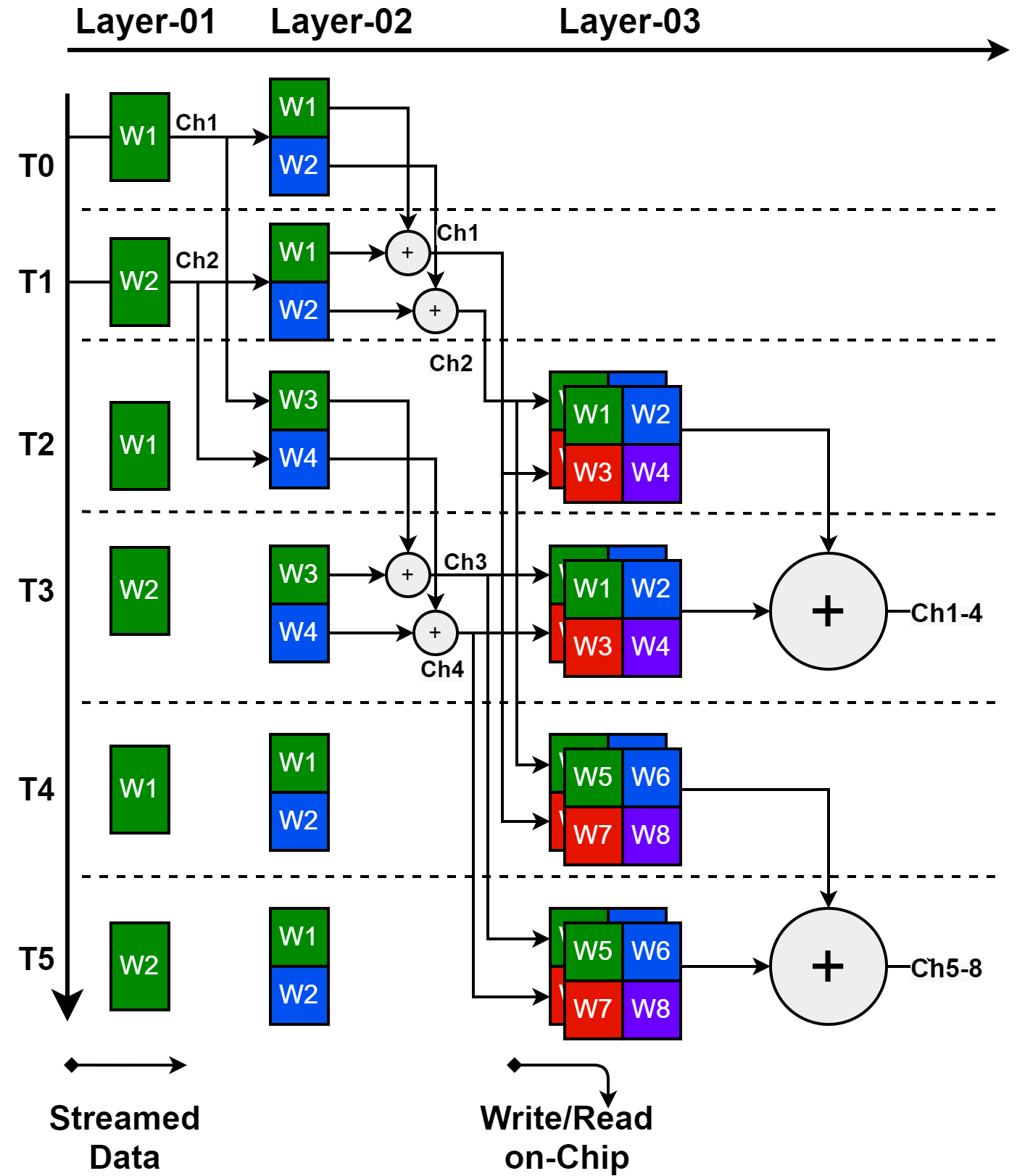}  
  \caption{Pipeline Scheduling for 1-2-4 design-space running 2-4-8 architecture.}
  \label{fig:Scheduling}
\end{figure}

Fig.~\ref{fig:Scheduling} illustrates how convolutional PEs are mapped onto the pipeline. Although pooling and non-linearity operations follow each convolutional layer, they are omitted from the figure for simplicity. The number of  \( \text{FC}_{\text{PEs}} \) can be scaled to parallelize FC computations.
Each active data channel streaming a feature map from the previous layer is assigned to a dedicated PE. However, to reduce resource overhead, the number of PEs executing the layer's filters can be limited by increasing each PE’s computational load. The goal of our optimization is to determine the optimal number of such dedicated PEs.

Latency for pipelined scheduling is modeled by Eq.~\eqref{eq:t_total} and Eq.~\eqref{eq:t_pipe}. The term $T_{\text{memory}}$ accounts for external memory access latency, which is negligible for the small- to medium-scale networks considered in this work.

\begin{equation}
T_{\text{total}} = T_{\text{pipe}} + T_{\text{memory}},
\label{eq:t_total}
\end{equation}

\begin{equation}
T_{\text{pipe}} = m \times P + (n - 1) \times I,
\label{eq:t_pipe}
\end{equation}
where, $m$ denotes the number of pipeline stages, $n$ is the number of input elements, $P$ is the system clock period, and $I$ is the initiation interval. For fully-pipelined hardware, the initiation interval is typically $I = P$, yielding one output per clock cycle after pipeline filling.

\begin{algorithm}[htbp]
\caption{MOGA NeuroForge}
\label{alg:moga_ode}
\begin{algorithmic}[1]
\State \textbf{Input:} CNN\_architecture $[p, k, ub, n]$, end\_criteria, constraints $[t, \text{DSP}, \text{LUT}, \text{BRAM}]$
\State \textbf{Output:} \texttt{ODE\_config}, $Y = \{Y_t, Y_{\text{DSP}}, Y_{\text{LUT}}, Y_{\text{BRAM}}\}$
\vspace{0.5em}
\State $l(i) = p(i) \times p(i{-}1)$ \Comment{$1 \leq p(i) \leq ub(i)$}
\State \texttt{ODE\_config} $\gets$ \texttt{Initialize}($l$)
\vspace{0.5em}
\While{end\_criteria is not satisfied}
    \State $i \gets i + 1$
    \State Select \texttt{ODE\_config}$(i)$ from parent pool \texttt{ODE\_config}$(g)$
    \State Perform crossover to generate offspring \texttt{ODE\_config}$(i+1)$
    \State Mutate \texttt{ODE\_config}$(i+1)$ using:
    \[
        x(i) \leftarrow 
        \begin{cases}
            x(i) - s(x(i) - lb(i)), & \text{if } t < r \\
            x(i) + s(ub(i) - x(i)), & \text{otherwise}
        \end{cases}
    \]
    \State Evaluate \texttt{ODE\_config}$(i+1)$:
    \For{$j = 1$ to $n$}
        \State $Y_{\text{DSP}} \mathrel{+}= l(j) \cdot k(j)^2 + l(n) \cdot 10$
        \State $Y_{\text{BRAM}} \mathrel{+}= p(j) \cdot k(j) - 1$
        \State Compute $Y_{\text{BRAM}}$ using Eq.~\eqref{eq:bram_est} \Comment{Refined BRAM estimate}
        \State Compute $Y_t$ using Eqs.~\eqref{eq:cpe_latency}, \eqref{eq:fc_reg}, \eqref{eq:fc_latency}, \eqref{eq:t_pipe}
        \State Lookup $Y_{\text{LUT}}$ from Table~\ref{tab:resource_utilization}
    \EndFor
    \State Apply constraints: \texttt{Constrain}(\texttt{ODE\_config}, $[t, \text{DSP}, \text{LUT}, \text{BRAM}]$)
\EndWhile
\vspace{0.5em}
\State \Return \texttt{ODE\_config}, $Y = \{Y_t, Y_{\text{DSP}}, Y_{\text{LUT}}, Y_{\text{BRAM}}\}$
\end{algorithmic}
\end{algorithm}

\begin{figure}[tb]
  \centering
  \includegraphics[width=0.6\columnwidth]{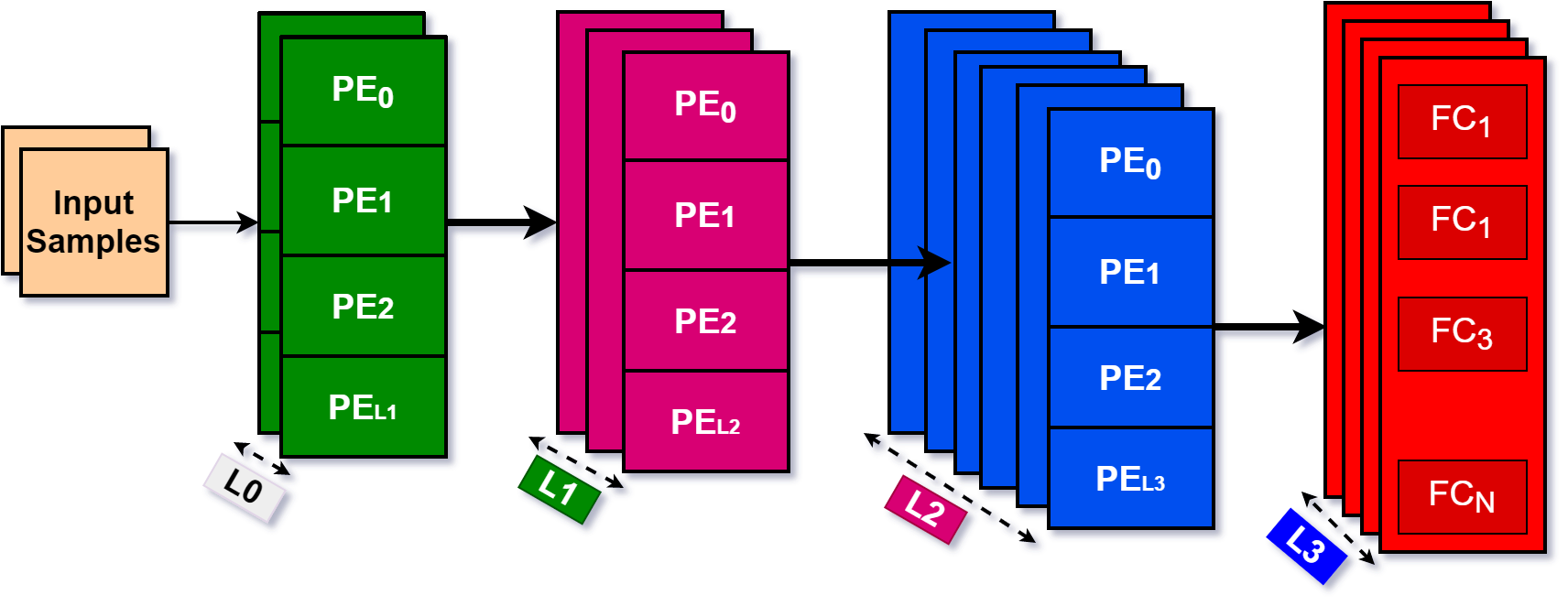}  
  \caption{Design-Space generations using (L0+L1+L2) × 4 CPEs.}
  \label{fig:DSE_ex}
\end{figure}

In Algorithm \ref{alg:moga_ode}, $t$ is the scaled distance of $x(i)$ from the $i$-th component of the lower bound, $lb(i)$. $s$ is a random variable drawn from a power distribution, and $r$ is a random number drawn from a uniform distribution.
The algorithm \ref{alg:moga_ode} takes three inputs: a fitness function, an input vector $\mathbf{P}$ of length $n$ (equal to the number of convolutional layers), and per-layer bounding constraints. Each element of $\mathbf{P}$ is constrained between 1 and $ub(i)$, the number of filters in layer $i$.

At each generation, the population of candidate solutions is evaluated using the fitness function, producing fitness scores that guide sampling for the next generation. The MOGA returns a set of Pareto-optimal solutions along with their objective vectors $\mathbf{Y}$, which include latency, and DSP, LUT, and BRAM usage.

Larger population sizes promote broader exploration of the design space, reducing the risk of missing global optima but increasing runtime. Accordingly, deeper networks are evaluated with larger populations. As MOGAs seek a diverse set of trade-offs rather than a single optimum, termination is based on a fixed number of generations or convergence criteria.

The DSP objective is estimated using the number of PEs required to fully parallelize computations in each layer, as defined by:

\begin{equation}
L(i) = P(i) \times P(i-1), \quad \text{with} \quad 1 \leq P(i) \leq ub(i),
\label{eq:dsp_pe}
\end{equation}

\begin{equation}
Y_{\text{DSP}} = \sum_{i=1}^{n} L(i) \times k(i)^2 \leq \frac{DSP_{\text{max}}}{k^2},
\label{eq:dsp_objective}
\end{equation}
where, $k(i)$ denotes the kernel size of layer $i$, and $\mathbf{L}$ is a vector of size $n$ where each element $L(i)$ corresponds to the number of PEs required for layer $i$. For example, if $P(1) = 3$ and $P(2) = 3$, then $L(2) = 9$, meaning 9 PEs are required to fully parallelize layer 2. Fig. \ref{fig:DSE_ex} provides a visual example.

The resource objectives $Y_{\text{DSP}}$, $Y_{\text{LUT}}$, and $Y_{\text{BRAM}}$, along with the latency estimate, are computed using the analytical models described in Section~2 and Eq.~\eqref{eq:dsp_pe} - \eqref{eq:dsp_objective}, and latency-related expressions defined earlier. These estimates depend on the number of PEs per convolutional layer and the degree of parallelism in the fully connected layer.

We use DSP slices as an optimizable objective along with latency in the examples below because, unlike LUT slices and BRAM, DSP slices are easier to predict as they are directly mapped for multiplications by the compiler. When DSP slices are used for optimizing resources against latency, the other metrics consistently satisfy the constraints and conditions set by the user as well.

\begin{table}[h!]
\centering
\caption{Resource utilization for different filter sizes}
\label{tab:resource_utilization}
\begin{tabular}{lcc cc}
\toprule
\multirow{2}{*}{\textbf{Filter Size}} 
& \multicolumn{2}{c}{\textbf{LUT Slices}} 
& \multicolumn{2}{c}{\textbf{Slice Registers}} \\
\cmidrule(lr){2-3} \cmidrule(lr){4-5}
& \textbf{Conv} & \textbf{Pool} & \textbf{Conv} & \textbf{Pool} \\
\midrule
2×2 & 550  & 300  & 1250  & 750  \\
3×3 & 850  & 420  & 2000  & 1000 \\
4×4 & 1400 & 700  & 3500  & 1400 \\
5×5 & 2000 & 900  & 5500  & 2200 \\
\bottomrule
\end{tabular}
\end{table}
\section{NeuroMorph: Online Design Reconfiguration}
\label{NeuroMorph}
Applying runtime reconfiguration to deep learning models presents a key challenge: although these models frequently reuse functional blocks, their parameters are highly interdependent and tightly coupled with the flow of data through the architecture. To address this, \textbf{NeuroMorph} introduces two reconfiguration techniques that preserve data integrity while enabling runtime performance trade-offs: \textit{depth-wise morphing} and \textit{width-wise morphing}.

\subsection{NeuroMorph Reconfiguration Strategies: Depth and Width}

To enable runtime adaptability and trade-offs between accuracy, power, and latency, \textbf{NeuroMorph} supports two morphing strategies for CNNs: \textit{depth-wise morphing} and \textit{width-wise morphing}. Both approaches restructure the network into modular subnetworks that can be selectively activated based on performance or resource constraints while maintaining the functional data flow.

\paragraph{Depth-Wise Morphing}
This strategy allows a CNN to operate using only portions of the full network by introducing intermediate output branches at various processing depths. Each segment—comprising a group of convolutional, non-linearity, and pooling layers—is treated as a standalone subnetwork, referred to as a \textit{Layer-Block}.

\begin{figure}[htbp]
\centering
\includegraphics[width=0.6\columnwidth]{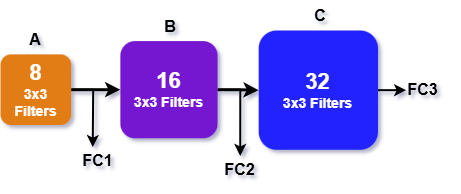}
\caption{MNIST architecture decomposed into three Layer-Blocks using depth-wise reconfiguration under the NeuroMorph framework.}
\label{fig:mnist_ODR}
\end{figure}

A complete network is composed of consecutive Layer-Blocks. For instance, consider a custom architecture \texttt{In-a-2a-3a-Out}. As illustrated in Fig.~\ref{fig:mnist_ODR}, Block A represents the first Layer-Block with $a$ filters, Block B the second with $2a$ filters, and Block C the final with $3a$ filters. The first subnetwork consists of Block A and a branch-off output layer. The second includes Blocks A and B with another output branch. The third configuration represents the original full network. Each subnetwork offers a distinct trade-off between accuracy and computational cost.

\paragraph{Width-Wise Morphing}
This technique reduces the number of active filters per convolutional layer at runtime—typically by half—to lower computational load. While \textit{depth-wise morphing} shortens the network by truncating layers, \textit{width-wise morphing} retains the full depth but scales down filter counts per layer.
For example, an MNIST model with layers of 8, 16, and 32 filters can be morphed at runtime to 4, 8, and 16 filters, producing a lightweight subnetwork with its own dedicated output head. Such morphs are particularly suited to edge devices operating under dynamic power or thermal constraints.
Width-wise morphing is also useful in distributed or multi-tenant environments where different devices may execute different slices of the model depending on their available compute. In all cases, this runtime flexibility allows for adaptive trade-offs between accuracy, energy, and latency without the need for re-synthesis or retraining.

\subsection{DistillCycle Training Methodology}

To support efficient and flexible inference across multiple configurations of a CNN, we introduce \textbf{DistillCycle Training}—a training strategy designed to jointly optimize both the full network and its smaller, reconfigurable subnetworks. Unlike traditional fine-tuning or one-shot pruning approaches, this method grows the network incrementally while ensuring all subnetworks remain accurate and deployable.

The training process follows three simple principles:
\begin{enumerate}
    \item \textbf{Grow progressively:} New layers are added step by step, starting from a shallow base network.
    \item \textbf{Train in cycles:} At each step, we alternate training between the current full network and its smaller subnetworks.
    \item \textbf{Use knowledge distillation:} Subnetworks are guided not only by labels, but also by predictions from the larger model, helping them retain performance with fewer parameters.
\end{enumerate}

At each stage, a subnetwork composed of the first $i$ layers is trained using a combined loss function. This includes both the usual classification loss and a distillation loss, which compares the subnetwork’s predictions to those of a larger, more accurate model (either pre-trained or trained in a previous step). This dual supervision ensures that every subnetwork remains valid, while the full model continues to improve with each added layer.

Training alternates between two optimization phases at each growth stage:

\paragraph{Teacher Phase (Full Network)} 
Train the current full model $\mathcal{N}_{\text{full}}^{(i)} = \mathcal{N}_i$ using standard supervised learning:
\begin{equation}
    \mathcal{L}_{\text{GT}} = \text{CrossEntropy}(y, \mathcal{N}_{\text{full}}^{(i)}(x)).
    \label{eq:gt_loss}
\end{equation}

\paragraph{Student Phase (Subnetwork with KD)} 
Train the current subnetwork $\mathcal{N}_i$ to match both the ground truth and the softened output of the full model (teacher). The Kullback-Leibler divergence is applied to softmax outputs scaled by a temperature factor $\tau$:
\begin{equation}
    \mathcal{L}_{\text{KD}} = \tau^2 \times \text{KL}\left(\sigma\left(\frac{x^{(t)}}{\tau}\right) \bigg\| \sigma\left(\frac{x^{(s)}}{\tau}\right)\right),
    \label{eq:kd_loss}
\end{equation}
\begin{equation}
    \mathcal{L}_{\text{total}} = \lambda \times \mathcal{L}_{\text{GT}} + (1 - \lambda) \times \mathcal{L}_{\text{KD}},
    \label{eq:total_loss}
\end{equation}

where, $x^{(t)}$ and $x^{(s)}$ denote the logits from the full and subnetwork models, respectively. The scalar $\lambda \in [0,1]$ balances the ground-truth supervision and the soft-label distillation. The temperature $\tau$ controls the sharpness of the softmax distribution.

\paragraph{Recursive Network Construction} 
The full network is built incrementally by appending a new Layer-Block $B_i$ at each stage:
\begin{equation}
    \mathcal{N}_{\text{full}}^{(i)} = \mathcal{N}_{\text{full}}^{(i-1)} \circ B_i
    \label{eq:recursive_network}
\end{equation}

\paragraph{Learning Rate Decay for Stability} 
To prevent catastrophic forgetting of early blocks during later stages, we apply exponentially decaying learning rates to layers $j < i$:
\begin{equation}
    \alpha_t^{(j)} = \alpha_0 \times \gamma^t \quad \text{with} \quad \gamma < 1.
    \label{eq:lr_decay}
\end{equation}

\paragraph{Overall Objective} 
The full DistillCycle training objective across $S$ incremental stages is:
\begin{equation}
    \min_{\theta} \sum_{i=1}^{S} \left[ \mathcal{L}_{\text{GT}}^{(i)} + \mathcal{L}_{\text{total}}^{(i)} \right].
    \label{eq:global_objective}
\end{equation}

This progressive expansion ensures that each subnetwork becomes a viable execution path, and the final full model is a composition of well-trained, resource-scalable modules. As a result, DistillCycle yields a family of runtime-ready configurations capable of dynamic inference under latency and energy constraints.
A complete description of the optimization loop is provided in Algorithm~\ref{alg:distillcycle}.

\begin{algorithm}[htbp]
\caption{DistillCycle Training (Depth- and Width-aware)}
\label{alg:distillcycle}
\begin{rmfamily}
\begin{algorithmic}[1]
\State \textbf{Input:} \texttt{dataset}, \texttt{CNN\_arch}, \texttt{morphing\_schedule}, \texttt{params} $[\alpha_0, \lambda, \tau, \gamma, \text{epochs}]$
\State \textbf{Output:} \texttt{Morphable\_Net}
\vspace{0.5em}
\State \texttt{net} $\gets$ \texttt{Initialize}(\texttt{CNN\_arch})
\State \texttt{trained\_paths} $\gets \emptyset$
\vspace{0.5em}

\For{$i = 1$ to $\text{size}(\texttt{morphing\_schedule})$}
    \State \texttt{config} $\gets$ \texttt{morphing\_schedule}[i]
    \State \texttt{subnet} $\gets$ \texttt{extract}(\texttt{net}, config.blocks, config.channels)
    \State $\alpha \gets \alpha_0$
    \vspace{0.5em}
    \For{$e = 1$ to \texttt{epochs}}
        \State \textbf{// Phase 1: Train Full Network (Teacher)} 
        \State \texttt{apply\_decay}(\texttt{net}, $\gamma^e$) \Comment{Apply Eq.~\eqref{eq:lr_decay}}
        \State Compute $\mathcal{L}_{\text{GT}}$ using Eq.~\eqref{eq:gt_loss}
        \State \texttt{net} $\gets$ \texttt{SGD}(\texttt{net}, $\mathcal{L}_{\text{GT}}, \alpha$)

        \State \textbf{// Phase 2: Train Subnetwork (Student) with KD}
        \For{$(x, y) \in \texttt{dataset}$}
            \State $x^{(t)} \gets \texttt{forward}(\texttt{net}, x)$
            \State $x^{(s)} \gets \texttt{forward}(\texttt{subnet}, x)$
            \State Compute $\mathcal{L}_{\text{KD}}$ using Eq.~\eqref{eq:kd_loss}
            \State Compute $\mathcal{L}_{\text{total}}$ using Eq.~\eqref{eq:total_loss}
            \State \texttt{subnet} $\gets$ \texttt{SGD}(\texttt{subnet}, $\mathcal{L}_{\text{total}}, \alpha$)
        \EndFor
        \State $\alpha \gets \alpha / 10$
    \EndFor

    \State \texttt{trained\_paths} $\gets$ \texttt{trained\_paths} $\cup$ \texttt{config}
    \State \texttt{net} $\gets$ \texttt{merge}(\texttt{subnet}, \texttt{net}) \Comment{Apply Eq.~\eqref{eq:recursive_network}}
\EndFor
\vspace{0.5em}
\State \Return \texttt{net with trained\_paths} \Comment{Final result optimizes Eq.~\eqref{eq:global_objective}}
\end{algorithmic}
\end{rmfamily}
\end{algorithm}

The proposed \textbf{DistillCycle Training} framework applies uniformly to both \textit{depth-wise} and \textit{width-wise morphing} strategies. While depth-wise morphing incrementally adds entire Layer-Blocks, width-wise morphing retains the network’s depth but progressively reduces the number of active filters in each convolutional layer. These reduced-width subnetworks serve as standalone execution paths, each equipped with a dedicated output head (e.g., fully connected layers) trained independently to match performance constraints at runtime.
Despite their structural differences, all subnetworks share a common input and perform the same classification task. This shared foundation allows early convolutional layers to learn broadly transferable features. The use of dedicated FC layers in each subnetwork serves to offset capacity loss, ensuring that each morphable path remains functionally valid even under severe resource constraints.
However, as the model grows deeper or more expressive, the training dynamics become increasingly constrained. The parameter landscape becomes harder to jointly optimize, especially when multiple subnetworks must maintain alignment with the full model. While DistillCycle Training improves subnetwork accuracy progressively (e.g., increasing from 76\% to 83.8\% in reduced-width configurations), the marginal gains to the full model's performance tend to plateau. This diminishing return is accompanied by rising training overhead, which scales with both the number of morphable configurations and the overall model complexity.
Consequently, deploying width-wise morphing effectively requires balancing training cost against runtime flexibility, and defining task-specific performance thresholds to ensure that reduced configurations remain viable under deployment conditions.

\section{Experimental Evaluation}
\label{expermental}

To evaluate \textbf{ForgeMorph}, we consider a range of hardware and runtime metrics across diverse architectures and datasets. For hardware generation with \textbf{NeuroForge}, we use pre-trained variants of LeNet-5~\cite{lecun1998} and custom architectures with varying depths, parameter counts, and computational costs, targeting MNIST~\cite{lecun1998}, CIFAR-10~\cite{krizhevsky2012}, and SVHN~\cite{netzer2011}. The selected benchmarks, summarized in Table~\ref{tab:architectures}, ensure fair comparison with existing literature.
Design Space Exploration (DSE) is driven by a Multi-Objective Genetic Algorithm (MOGA), which searches for Pareto-optimal hardware mappings balancing area and latency under resource and timing constraints~\cite{cardoso2017}. We validate DSE effectiveness by implementing 22 randomly selected configurations on FPGA and comparing measured results against model estimates.
For runtime reconfiguration via \textbf{NeuroMorph}, we assess accuracy, latency, and power trade-offs before and after dynamic mode switching. Modular network designs, structured as $a$–$2a$–$3a$–$4a$ convolutional pipelines, provide a clear space for both depth-wise and width-wise reconfiguration, treating each convolutional layer as an independent subnetwork. This setup enables flexible runtime scaling while preserving data integrity and minimizing accuracy loss.

In addition, to benchmark against other compilers and edge platforms, we adapted \textit{ForgeMorph} to support MobileNetV2, ResNet-50, SqueezeNet, and YOLOv5-Large models, demonstrating competitive results on ImageNet and COCO datasets through minor custom modifications to the compiler.

\textbf{ForgeMorph} uses a streaming interface optimized for real-time, sample-by-sample processing, unlike traditional array-based PE architectures designed for offline inference. The network is compiled into a pipelined architecture that eliminates the need for full-frame reads, enabling immediate data flow.
Rather than requiring static configuration of image dimensions or blanking intervals, our pipeline uses a centralized parameter update during serialization. Block-level parameters (e.g., line buffer sizes) are automatically adjusted by the compiler based on the input size, avoiding manual reconfiguration.
Although parameters are fixed at generation time, control signals maintain synchronization and ensure correct segmentation during streaming, regardless of the data source.
All designs are generated with \textit{NeuroForge} and synthesized using Xilinx Vivado. After place-and-route, SAIF files are used for accurate power estimation. \textit{NeuroMorph} handles runtime reconfiguration via clock gating \cite{clk_gating}, selectively disabling inactive layers/channels to minimize power. These blocks resume execution only after reactivation and a full-frame delay. All results are reported on a Xilinx Zynq-7100 operating at 250,MHz.

\begin{table}[h!]
\centering
\caption{Architectures used for validation.}
\label{tab:architectures}
\begin{tabular}{lccc}
\toprule
\textbf{Dataset} & \textbf{Architecture} & \textbf{\# Parameters} & \textbf{\# Operations} \\
\midrule
MNIST       & 8-16-32          & 333.72K   & 6.79M \\
SVHN        & 8-16-32-64       & 639.58K   & 32.2M \\
CIFAR-10    & 8-16-32-64-64    & 676K      & 83M \\
ImageNet    & ResNet-50        & 25.56M    & 4.1B \\
ImageNet    & MobileNetV2      & 2.26M     & 300M \\
ImageNet    & SqueezeNet       & 1.24M     & 833M \\
COCO 2017   & YOLOv5-Large     & 46.5M     & 154.0B \\
\bottomrule
\end{tabular}
\end{table}

\subsection{NeuroForge: Offline Design Space Exploration Results}

\begin{figure}[htbp]
  \centering
  \includegraphics[width=0.9\columnwidth]{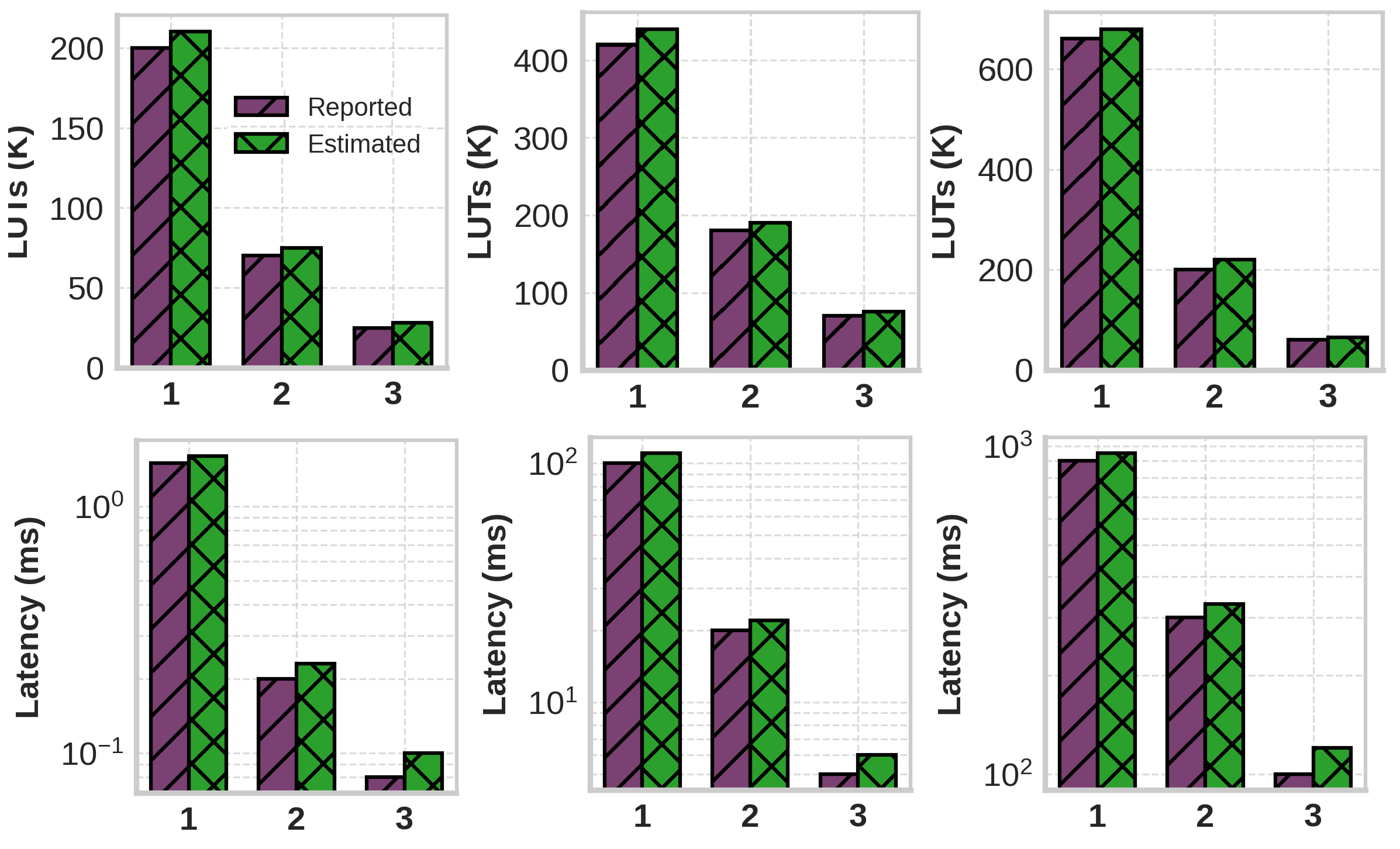}  
  \caption{Estimated and reported resources and latencies for 4 different architectures (MNIST 8-16-32, SVHN 8-16-32-64, CIFAR 8-16-32-64-64) deployed in 3 NeuroForge configurations of varying sizes.}
  \label{fig:ODE_resource_vlidation}
\end{figure}

\begin{table*}[h!]
\centering
\caption{
Estimated and reported resource usage for FPGA deployments of CNN architectures generated by NeuroForge across three datasets: MNIST (8-16-32), SVHN (8-16-32-64), and CIFAR-10 (8-16-32-64-64). Each row corresponds to a distinct NeuroForge configuration, with comparisons between real measurements and analytical predictions. 
}

\label{tab:merged_resource_usage}
\begin{tabular}{l lccc ccc ccc cc c}
\toprule
\textbf{Dataset} & \textbf{Design PEs} 
& \multicolumn{3}{c}{\textbf{DSP Slices}} 
& \multicolumn{3}{c}{\textbf{LUT Slices}} 
& \multicolumn{3}{c}{\textbf{Block RAM}} 
& \multicolumn{2}{c}{\textbf{Latency (ms)}} 
& \textbf{Power (mW)} \\
\cmidrule(lr){3-5} \cmidrule(lr){6-8} \cmidrule(lr){9-11} \cmidrule(lr){12-13}
& & Real & MOGA & Err.\% & Real & MOGA & Err.\% & Real & MOGA & Err.\% & MOGA & Real & \\
\midrule
\rowcolor{orange!15}
MNIST & 648 & 6000 & 6410 & 2.44\% & 657K & 641K & 2.44\% & 1325 & 1325 & 0\% & 0.010 & NA & NA \\
\rowcolor{green!15}
MNIST & 164 & 1556 & 1556 & 0\% & 192K & 200.56K & 4.27\% & 356 & 356 & 0\% & 0.041 & 0.042 & 743 \\
\rowcolor{green!15}
MNIST & 42  & 485  & 485  & 0\% & 66K  & 68.28K  & 3.39\% & 98  & 98  & 0\% & 0.164 & 0.165 & 660 \\
\rowcolor{green!15}
MNIST & 11  & 179  & 179  & 0\% & 24K  & 26.14K  & 8.18\% & 29  & 29  & 0\% & 0.660 & 0.669 & 578 \\
\rowcolor{green!15}
MNIST & 3   & 35   & 35   & 0\% & 6.59K & 7.26K   & 4.96\% & 9   & 9   & 0\% & 3.920 & 4.000 & 475 \\

\rowcolor{red!15}
SVHN & 2702 & 24000 & 24000 & 0\% & 1750K & 2000K & 12.50\% & 5000 & 5000 & 0\% & 0.012 & NA & NA \\
\rowcolor{orange!15}
SVHN & 684  & 6000  & 6000  & 0\% & 657K  & 685K  & 4.09\% & 1428 & 1428 & 0\% & 0.256 & NA & NA \\
\rowcolor{green!15}
SVHN & 196  & 1924  & 1924  & 0.00\% & 215K  & 227K  & 5.28\% & 414  & 414  & 0\% & 1.390 & 1.720 & 824 \\
\rowcolor{green!15}
SVHN & 45   & 485   & 485   & 0\% & 69K   & 71K   & 2.82\% & 105  & 105  & 0\% & 8.890 & 12.640 & 711 \\
\rowcolor{green!15}
SVHN & 4    & 37    & 37    & 0\% & 8K    & 8.5K  & 5.88\% & 12   & 12   & 0\% & 95.120 & 123.620 & 692 \\

\rowcolor{red!15}
CIFAR-10 & 2840 & 25000 & 25000 & 0\% & 1780K & 2000K & 11.00\% & 6000  & 6000  & 0\% & 0.288 & NA & NA \\
\rowcolor{orange!15}
CIFAR-10 & 430  & 4000  & 4000  & 0\% & 408K  & 425K  & 4.00\% & 906   & 906   & 0\% & 10.80 & NA & NA \\
\rowcolor{green!15}
CIFAR-10 & 109  & 1061  & 1061  & 0\% & 119K  & 125K  & 4.80\% & 241   & 241   & 0\% & 260.0 & 277.3 & 1530 \\
\rowcolor{green!15}
CIFAR-10 & 76   & 724   & 724   & 0\% & 78K   & 83K   & 6.02\% & 164   & 164   & 0\% & 91.11  & 113.0 & 1950 \\
\rowcolor{green!15}
CIFAR-10 & 22   & 218   & 218   & 0\% & 27K   & 27.9K & 3.25\% & 54    & 54    & 0\% & 1315 & 1427 & 1461 \\
\rowcolor{green!15}
CIFAR-10 & 1    & 46    & 46    & 0\% & 39K   & 42K   & 7.14\% & 15    & 15    & 0\% & 1723 & 1835 & 1121 \\
\bottomrule
\end{tabular}
\end{table*}

\begin{table*}[h!]
\centering
\captionsetup{justification=justified}  
\caption{
Comparison of FPGA implementations using different compilers on ImageNet (classification) and COCO 2017 (detection). 
\textit{NeuroMorph} applies depth-wise reconfiguration to produce two subnetworks where possible. 
}

\label{tab:fpga_comparison_final}
\renewcommand{\arraystretch}{0.8}
\begin{tabular}{lcccccc}
\toprule
\textbf{Framework} & \textbf{Precision} & 
\textbf{Throughput (FPS) $\uparrow$} & 
\textbf{Top-1 Acc (\%) $\uparrow$} & 
\textbf{Energy (J/frame) $\downarrow$} & 
\textbf{Freq (MHz)} & 
\textbf{FPGA} \\
\midrule
\multicolumn{7}{c}{\textbf{MobileNetV2 (ImageNet)}} \\
\midrule
\textbf{NeuroForge-16} & int16 & 381.3 & \textbf{75.1} & 0.35 & 250 & Zynq-7100 \\
\textbf{NeuroForge-8}  & int8  & \textbf{785.0} & 73.0 & 0.22 & 250 & Zynq-7100 \\
\textbf{NeuroMorph (full/split)} & int8 & 765.0 / \textbf{1527.4} & 70.5 / 68.0 & 0.21 / \textbf{0.15} & 250 & Zynq-7100 \\
Vitis AI \cite{vitis_ai}        & int8 & 765.0 & 73.5 & 0.20 & 300 & ZCU102 \\
hls4ml \cite{hls4ml}          & int8 & 815.7 & 73.1 & 0.19 & 200 & Kintex-7 \\
TVM   \cite{tvm}           & int8 & NA & NA & NA & NA & NA \\
OpenVINO \cite{openvino}         & int8 & 300.0 & 71.8 & NA & 300 & Arria 10 GX 660 \\
\midrule
\multicolumn{7}{c}{\textbf{ResNet-50 (ImageNet)}} \\
\midrule
\textbf{NeuroForge-16} & int16 & 113.1 & \textbf{77.2} & 0.75 & 250 & Zynq-7100 \\
\textbf{NeuroForge-8}  & int8  & \textbf{225.0} & 76.3 & 0.48 & 250 & Zynq-7100 \\
\textbf{NeuroMorph (full/split)} & int8 & 215.5 / \textbf{448.1} & 74.0 / 71.8 & 0.47 / \textbf{0.35} & 250 & Zynq-7100 \\

Vitis AI \cite{vitis_ai}      & int8 & 214.0 & 76.5 & 0.89 & 300 & ZCU102 \\
hls4ml \cite{hls4ml}         & int8 & 267.9 & 76.2 & 0.40 & 200 & Kintex-7 \\
TVM   \cite{tvm}            & int8 & 102.5 & 74.4 & NA & 200 & ZCU102 \\
OpenVINO \cite{openvino}         & int8 & 132.3 & 75.5 & NA & 300 & Arria 10 GX 660 \\
\midrule
\multicolumn{7}{c}{\textbf{SqueezeNet (ImageNet)}} \\
\midrule
\textbf{NeuroForge-16} & int16 & 728.9 & \textbf{60.1} & 0.18 & 250 & Zynq-7100 \\
\textbf{NeuroForge-8}  & int8  & \textbf{1615.0} & 58.9 & 0.14 & 250 & Zynq-7100 \\
\textbf{NeuroMorph (full/split)} & int8 & 1580.0 / \textbf{2943.1} & 56.7 / 55.0 & 0.13 / \textbf{0.09} & 250 & Zynq-7100 \\
Vitis AI \cite{vitis_ai}       & int8 & 1527.0 & 59.3 & 0.16 & 300 & ZCU102 \\
hls4ml \cite{hls4ml}           & int8 & 1610.0 & 59.0 & 0.13 & 200 & Kintex-7 \\
TVM   \cite{tvm}             & int8 & 497.5 & 59.2 & NA & NA & NA \\
OpenVINO \cite{openvino}        & int8 & NA & NA & NA & NA & NA \\
\midrule
\multicolumn{7}{c}{\textbf{YOLOv5-Large (COCO 2017)}} \\
\midrule
\textbf{NeuroForge-16} & int16 & 97.7 & \textbf{62.4} & 1.20 & 250 & Zynq-7100 \\
\textbf{NeuroForge-8}  & int8  & \textbf{215.0} & 60.3 & 0.80 & 250 & Zynq-7100 \\
Vitis AI \cite{vitis_ai}       & int8 & 202.0 & 60.8 & 0.75 & 300 & ZCU102 \\
hls4ml \cite{hls4ml}           & int8 & NA & NA & NA & NA & NA \\
TVM   \cite{tvm}             & int8 & 123.4 & 60.5 & NA & NA & NA \\
OpenVINO \cite{openvino}       & int8 & 140.0 & 61.0 & NA & 300 & Arria 10 GX 660 \\
\bottomrule
\end{tabular}
\end{table*}

\begin{table*}[htbp]
\centering
\caption{Resource Utilization Comparison across models. Values are reported after place and route on Zynq-7100 (444K LUTs, 26.5Mb BRAM, 2020 DSPs). Percentage values indicate total device utilization.}
\label{tab:resource_utilization_zynq7100}
\renewcommand{\arraystretch}{0.95}
\setlength{\tabcolsep}{9pt}
\begin{tabular}{llcccccc}
\toprule
\textbf{Model} & \textbf{Precision} & \textbf{kLUT (\%)} & \textbf{BRAM (Mb, \%)} & \textbf{FF (K, \%)} & \textbf{DSP (\%)} & \textbf{Freq. (MHz)}  \\
\midrule
\multicolumn{7}{c}{\textbf{MobileNetV2 (ImageNet)}} \\
\midrule
NeuroForge & int16 & 122.5 (27.6\%) & 18.2 (68.7\%) & 135.0 (30.4\%) & 1638 (81.1\%) & 250  \\
NeuroForge & int8  & 103.6 (23.3\%) & 15.6 (58.9\%) & 119.4 (26.9\%) & 1415 (70.0\%) & 250  \\
\midrule
\multicolumn{7}{c}{\textbf{ResNet-50 (ImageNet)}} \\
\midrule
NeuroForge & int16 & 135.3 (30.5\%) & 19.6 (74.0\%) & 152.2 (34.3\%) & 1710 (84.7\%) & 250  \\
NeuroForge & int8  & 116.7 (26.3\%) & 16.9 (63.8\%) & 137.0 (30.9\%) & 1532 (75.9\%) & 250  \\
\midrule
\multicolumn{7}{c}{\textbf{SqueezeNet (ImageNet)}} \\
\midrule
NeuroForge & int16 & 88.4 (19.9\%) & 12.3 (46.4\%) & 102.1 (23.0\%) & 1120 (55.4\%) & 250  \\
NeuroForge & int8  & 75.7 (17.1\%) & 10.1 (38.1\%) & 91.5 (20.6\%)  & 987 (48.9\%)  & 250  \\
\midrule
\multicolumn{7}{c}{\textbf{YOLOv5-Large (COCO 2017)}} \\
\midrule
NeuroForge & int16 & 210.1 (47.3\%) & 24.5 (92.5\%) & 187.6 (42.3\%) & 1942 (96.1\%) & 250  \\
NeuroForge & int8  & 185.8 (41.8\%) & 21.7 (81.9\%) & 165.3 (37.2\%) & 1760 (87.1\%) & 250  \\
\bottomrule
\end{tabular}
\end{table*}

\begin{table*}[htbp]
\centering
\caption{Comparison of Edge Devices on Latency, Power, and Inference Efficiency on MobileNetV1 (based on public benchmarks from MLPerf~\cite{mlperf}).}
\label{tab:device_comparison}
\renewcommand{\arraystretch}{1.3}
\setlength{\tabcolsep}{2.5pt}
\begin{tabular}{lccccccccccc}
\toprule
\textbf{Metric} &
\textbf{RasPi4~\cite{raspi4}} &
\textbf{NCS~\cite{movidius1}} &
\textbf{NCS2~\cite{movidius2}} &
\textbf{Nano~\cite{nano}} &
\textbf{TX2~\cite{tx2}} &
\textbf{XavNX~\cite{xaviernx}} &
\textbf{AGX~\cite{xavier}} &
\textbf{Tinker~\cite{tinker}} &
\textbf{Coral~\cite{coral}} &
\textbf{Snap888~\cite{snapdragon}} &
\textbf{FPGA~(ours)} \\
\midrule
\textbf{Latency (ms)} & 
480.3 & 
115.7 & 
87.2 & 
72.3 & 
9.17 & 
0.95 & 
\textbf{0.53} & 
14.6 & 
15.7 & 
11.6 & 
3.72 \\
\textbf{Power (W)} & 
1.3 & 
2.5 & 
1.5 & 
10.0 & 
15.0 & 
20.0 & 
30.0 & 
7.8 & 
5.0 & 
5.0 & 
\textbf{1.53} \\
\textbf{Inference/Watt} & 
1.6 & 
3.5 & 
7.6 & 
1.4 & 
7.3 & 
52.6 & 
\textbf{62.9} & 
8.8 & 
12.7 & 
17.2 & 
\textbf{178} \\
\bottomrule
\end{tabular}
\end{table*}

We begin by validating the accuracy of \textbf{NeuroForge}’s analytical estimators, which guide the design space exploration process. Fig.~\ref{fig:ODE_results} illustrates the Pareto front discovered by the Multi-Objective Genetic Algorithm for the CIFAR-10 8-16-32-64-64 model, revealing efficient trade-offs between DSP slices and latency. Most feasible designs cluster near the Pareto front, allowing the compiler to prioritize high-efficiency mappings while discarding suboptimal configurations with diminishing returns.

Fig.~\ref{fig:ODE_resource_vlidation} compares estimated and post-synthesis values for latency and resource usage across three configurations for four networks. NeuroForge achieves over 95\% accuracy for DSP and BRAM utilization estimates. As expected, LUT usage is less accurate due to unmodeled control logic and routing overheads, with the largest deviation observed in the most complex design. Latency predictions are typically within 10–15\% of actual measurements, with discrepancies primarily due to memory and control overheads excluded from the model. These results demonstrate that NeuroForge can perform fast and reliable estimation—sufficiently accurate to guide hardware search at scale.

To evaluate how well NeuroForge spans the design space under deployment constraints, Table~\ref{tab:merged_resource_usage} reports estimated and post-synthesis utilization across three models. The \textit{Design PEs} column reflects the number of processing elements allocated per configuration, which generally correlates with throughput. Because pipelining and resource reuse create multiple valid schedules for a given PE count, this value is an abstracted design indicator—only one such mapping lies on the Pareto front, and all entries shown are Pareto-optimal.
Color coding indicates deployment feasibility on the Zynq-7100: green for fully mappable designs, orange for configurations requiring runtime switching, and red for infeasible mappings. The coverage of feasible and near-feasible designs confirms that NeuroForge produces solutions across a range of resource budgets, supporting both fixed and reconfigurable deployment via \textit{NeuroMorph}.

Next, we compare ForgeMorph to state-of-the-art FPGA compilers on throughput, accuracy, and energy. Table~\ref{tab:fpga_comparison_final} show that NeuroForge-generated static designs match or exceed leading tools such as Vitis AI and hls4ml. For example, on ResNet-50, \textit{NeuroForge-8} achieves 225 FPS (vs. 214 FPS for Vitis AI) while consuming 0.48 J/frame (vs. 0.89 J/frame). Use of INT16 precision provides higher accuracy when required, consistently delivering the top score across all networks.
\textit{NeuroMorph}'s reconfigurable variant further improves performance. On SqueezeNet, it delivers 2943 FPS with only a 5\% accuracy drop—significantly outperforming static baselines. Table~\ref{tab:resource_utilization_zynq7100} highlights how NeuroForge mappings span a broad range of resource footprints, allowing users to select configurations aligned with specific latency, energy, or area constraints.
Critically, ForgeMorph is the only evaluated compiler supporting runtime reconfiguration. On MobileNetV2 and ResNet-50, \textit{NeuroMorph} doubles throughput (1527 vs. 765 FPS) and reduces energy by more than 25\% (0.15 vs. 0.21 J/frame), all without retraining or re-synthesis. This runtime flexibility is absent from all other frameworks.

Finally, Table~\ref{tab:device_comparison} compares edge platform efficiency. On Zynq-7100, ForgeMorph achieves 3.72 ms latency and 1.53 W power, resulting in 178 inferences/Watt—over 2.8× higher than the next-best system (AGX Xavier).

\begin{figure}[htbp]
  \centering
  \includegraphics[width=0.9\columnwidth]{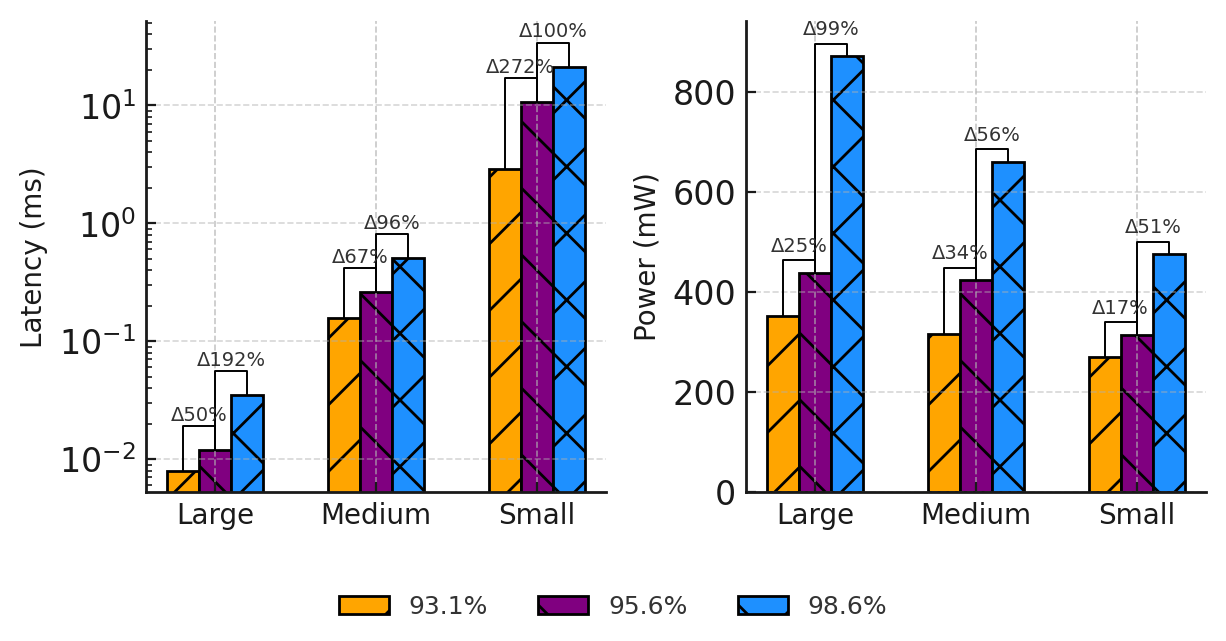}  
  \caption{Depth-wise Reconfiguration on MNIST-8-16-32 Using \textit{NeuroMorph}. Latency, power and accuracy are reported for both full and NueroMorph reconfigured deployment.}
  \label{fig:depth_wise_bar}
\end{figure}

\begin{figure*}[htbp]
  \centering
  \includegraphics[width=0.8\textwidth]{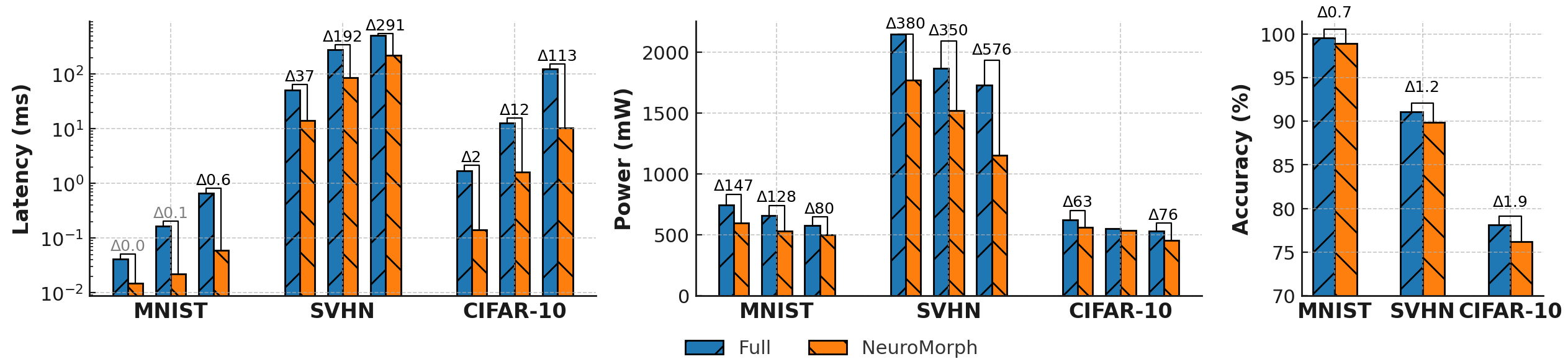}  
  \caption{Width-wise Reconfiguration Using \textit{NeuroMorph}. Latency, power and accuracy are reported for both full and NueroMorph reconfigured deployment.}
  \label{fig:width_wise_bar}
\end{figure*}

\subsection{NeuroMorph: Online Design Reconfiguration Results}
We now evaluate \textit{NeuroMorph}'s runtime behavior through FPGA-based experiments using both depth-wise and width-wise reconfiguration. These tests validate how the system responds to latency, power, and accuracy trade-offs when activating different subnetworks on-chip, based on configurations generated by \textit{NeuroForge}.
\subsubsection{Depth-wise Morphing}
We begin with depth-wise reconfiguration experiments to quantify the impact of activating partial depthwise network blocks at runtime. Figure~\ref{fig:depth_wise_bar} presents trade-offs achieved by activating three reconfigurable subnets derived from three \textit{NeuroForge}-generated configurations. All selected configurations lie on the Pareto front, spanning a range of resource-performance trade-offs. Activating smaller subnets leads to latency reductions of up to 200\% and power savings exceeding 90\%, with accuracy drops limited to 5.5\%. This demonstrates \textit{NeuroMorph}'s ability to modulate execution cost on the fly, based on application needs—without retraining or redeployment.

\subsubsection{Width-wise Morphing}

We next evaluate runtime reconfiguration across width-wise variants using three \textit{NeuroForge} configurations per dataset, each with different PE allocations. Fig.~\ref{fig:width_wise_bar} presents results for MNIST, SVHN, and CIFAR-10, where \textit{NeuroMorph} selectively activates a subset of filters to reduce the active compute footprint during inference.

Significant performance gains are observed: latency drops by up to 91\% on MNIST and 84\% on SVHN, while power consumption decreases by over 300\,mW in deeper models. Accuracy degradation remains under 2\% across all configurations. These results validate that \textit{NeuroMorph} enables low-power, high-efficiency execution with minimal impact on task accuracy.

Together, these results confirm that combining \textit{NeuroForge} and \textit{NeuroMorph} yields a highly flexible deployment stack. Developers can pre-compile multiple performance modes into a single bitstream and switch between them at runtime to accommodate dynamic power or latency constraints.
\textit{Depth-wise morphing} disables entire layer blocks, yielding substantial latency reductions—up to 8.3× speedup—by shortening the streaming pipeline and eliminating both compute and memory operations for deactivated layers. This makes it particularly effective for energy-constrained settings where throughput is secondary. In contrast, \textit{width-wise morphing} retains the network depth while reducing the number of active channels per layer. Although latency gains are more moderate (up to 4.5×), this mode preserves inter-layer structure, resulting in higher accuracy retention and more stable predictions—making it more suitable for latency-sensitive tasks.
From a hardware mapping perspective, these behaviours stem directly from NeuroForge’s streaming pipeline with statically allocated PEs. Depth-wise morphing disables pipeline stages entirely, yielding sharper drops in execution time, especially in designs with a low PE budget where each stage constitutes a bottleneck. Width-wise morphing, on the other hand, scales down intra-stage workload, achieving predictable speedups while maintaining alignment with the original architecture.
Overall, latency, compute, and memory savings can vary by up to 100× between configurations, while runtime power reductions exceed 50\% without retraining. This level of reconfigurability makes ForgeMorph particularly valuable for edge and mission-critical deployments, where system conditions shift unpredictably, and flexible adaptation is essential.

\section{Acknowledgment}
This research was funded by the European Union’s HORIZON Research and Innovation Programme under grant agreement No 101120657, project ENFIELD (European Lighthouse to Manifest Trustworthy and Green AI).

\section{Conclusion and Discussion}
\label{sec:conclusions}
In this paper, we present an automated methodology for hardware model and code generation targeting CNNs. Our compiler, \textit{ForgeMorph}, produces multiple hardware configurations for a given network, enabling performance-aware operation modes driven by user-defined latency and resource constraints. Leveraging analytical models for latency and resource usage, the generated designs achieve throughput-resource trade-offs of up to 95$\times$, 71$\times$, and 18$\times$ for MNIST, CIFAR-10, and SVHN, respectively.
At the heart of this approach is \textit{NeuroForge}, a design-time exploration engine based on a Multi-Objective Genetic Algorithm that systematically navigates the device-specific configuration space. Complementing this, \textit{NeuroMorph} enables runtime performance tuning through dynamic reconfiguration, allowing users to trade accuracy for reduced power and latency. This is achieved via a dual-path training strategy, which allows a single CNN to support multiple execution paths with shared weights and minimal accuracy loss. NeuroMorph can achieve runtime power reductions of up to 90\%, and latency improvements of 50$\times$.
Future work includes automating NeuroMorph's configuration extraction via combinatorial analysis, enabling automatic selection of optimal runtime paths that meet application-specific accuracy constraints. Another promising direction is extending support to non-traditional CNN architectures such as Transformers. 
The proposed workflow is validated using pre-trained and reference models. While optimality within the device-specific layout space is not guaranteed, the genetic search strategy—combined with constraint filtering—effectively converges toward Pareto-optimal solutions. Future work may explore local optimality assessment and layout-aware design exploration.
Regarding portability, tools such as MyHDL \cite{decaluwe2004} and PyHDL \cite{haglund2003} provide Python-based alternatives to VHDL.

\bibliographystyle{IEEEtran}  

\begin{thebibliography}{99} 
\bibitem{stimpson2017} A. J. Stimpson, M. B. Tucker, M. Ono, A. Steffy, and M. L. Cummings, "Modeling risk perception for mars rover supervisory control: Before and after wheel damage," in \textit{Aerospace Conference, 2017 IEEE}, Montana. USA, Mar 4 2017: IEEE, pp. 1-8.

\bibitem{mazouz2019} A. Mazouz and C. P. Bridges, "Multi-Sensory CNN Models for Close Proximity Satellite Operations," in \textit{2019 IEEE Aerospace Conference, 2-9 March 2019}, 2019, pp. 1-7, doi: 10.1109/AERO.2019.8741723.

\bibitem{AEM2024} Mazouz, Alaa Eddine, and Van-Tam Nguyen. "Online continual streaming learning for embedded space applications." Journal of Real-Time Image Processing 21.3 (2024): 68.

\bibitem{Liu2024} Liu, Jun, et al. "TSLA: A Task-Specific Learning Adaptation for Semantic Segmentation on Autonomous Vehicles Platform." IEEE Transactions on Computer-Aided Design of Integrated Circuits and Systems (2024).

\bibitem{kasem2020} A. Kasem, A. Bouzid, A. Reda, and J. Vásárhelyi, "A Survey about Intelligent Solutions for Autonomous Vehicles based on FPGA," \textit{Carpathian Journal of Electronic and Computer Engineering}, vol. 13, pp. 9-13, 12/31 2020, doi: 10.2478/cjece-2020-0XXX.

\bibitem{kouris2019} A. Kouris, S. Venieris, and C. Bouganis, Towards Efficient On-Board Deployment of DNNs on Intelligent Autonomous Systems. 2019, pp. 568-573.

\bibitem{Chai2023} Chai, Zhuomin, et al. "Circuitnet: An open-source dataset for machine learning in vlsi cad applications with improved domain-specific evaluation metric and learning strategies." IEEE Transactions on Computer-Aided Design of Integrated Circuits and Systems 42.12 (2023): 5034-5047.

\bibitem{santos2020} D. Almeida dos Santos, D. Zolett, M. Belli, F. Viel, and C. Zeferino, An Analysis of the Implementation of Edge Detection Operators in FPGA. 2020, pp. 163-167.

\bibitem{Habeeb2023} Habeeb, P., et al. "Verification of camera-based autonomous systems." IEEE Transactions on Computer-Aided Design of Integrated Circuits and Systems 42.10 (2023): 3450-3463.

\bibitem{yih2018} M. Yih, J. Ota, J. Owens, and P. Muyan-Ozcelik, FPGA versus GPU for Speed-Limit-Sign Recognition. 2018, pp. 843-850.

\bibitem{Han2024} Han, Lixia, et al. "CoMN: Algorithm-Hardware Co-Design Platform for Nonvolatile Memory-Based Convolutional Neural Network Accelerators." IEEE Transactions on Computer-Aided Design of Integrated Circuits and Systems 43.7 (2024): 2043-2056.

\bibitem{Yaping2019}Li, Yaping, et al. "An artificial neural network assisted optimization system for analog design space exploration." IEEE Transactions on Computer-Aided Design of Integrated Circuits and Systems 39.10 (2019): 2640-2653.

\bibitem{venkatesh2017} G. Venkatesh, E. Nurvitadhi, and D. Marr, "Accelerating Deep Convolutional Networks using low-precision and sparsity," in \textit{2017 IEEE International Conference on Acoustics, Speech and Signal Processing (ICASSP)}, 5-9 March 2017 2017, pp. 2861-2865, doi: 10.1109/ICASSP.2017.7952679.

\bibitem{nurvitadhi2017} E. Nurvitadhi et al., "Can FPGAs Beat GPUs in Accelerating Next-Generation Deep Neural Networks?," presented at the \textit{Proceedings of the 2017 ACM/SIGDA International Symposium on Field-Programmable Gate Arrays}, Monterey, California, USA, 22 Feb, 2017.

\bibitem{gan2016} F. Gan, H. Zuyi, C. Song, and W. Feng, "Energy-efficient and high-throughput FPGA-based accelerator for Convolutional Neural Networks," in \textit{2016 13th IEEE International Conference on Solid-State and Integrated Circuit Technology (ICSICT)}, 25-28 Oct. 2016 2016, pp. 624-626, doi: 10.1109/ICSICT.2016.7998996.


\bibitem{DNNBuilder}
Zhang, Xiaofan, et al. "DNNBuilder: An automated tool for building high-performance DNN hardware accelerators for FPGAs." 2018 IEEE/ACM International Conference on Computer-Aided Design (ICCAD). IEEE, 2018.

\bibitem{AutoDNN}
Xu, Pengfei, et al. "AutoDNNchip: An automated DNN chip predictor and builder for both FPGAs and ASICs." Proceedings of the 2020 ACM/SIGDA International Symposium on Field-Programmable Gate Arrays. 2020.

\bibitem{DeepBurning}
Cai, Xuyi, et al. "Deepburning-seg: Generating dnn accelerators of segment-grained pipeline architecture." 2022 55th IEEE/ACM International Symposium on Microarchitecture (MICRO). IEEE, 2022.
\bibitem{DNNExplorer}
Zhang, Xiaofan, et al. "DNNExplorer: a framework for modeling and exploring a novel paradigm of FPGA-based DNN accelerator." Proceedings of the 39th International Conference on Computer-Aided Design. 2020.

\bibitem{FINN-R}
Blott, Michaela, et al. "FINN-R: An end-to-end deep-learning framework for fast exploration of quantized neural networks." ACM Transactions on Reconfigurable Technology and Systems (TRETS) 11.3 (2018): 1-23.

\bibitem{kouris2018} A. Kouris, S. Venieris, and C. Bouganis, CascadeCNN: Pushing the Performance Limits of Quantisation in Convolutional Neural Networks. 2018.

\bibitem{venieris2016fccm} S. I. Venieris and C. Bouganis, "fpgaConvNet: A Framework for Mapping Convolutional Neural Networks on FPGAs," in \textit{2016 IEEE 24th Annual International Symposium on Field-Programmable Custom Computing Machines (FCCM)}, 1-3 May 2016 2016, pp. 40-47, doi: 10.1109/FCCM.2016.22.

\bibitem{venieris2018tnnls} S. Venieris and C. Bouganis, "fpgaConvNet: Mapping Regular and Irregular Convolutional Neural Networks on FPGAs," \textit{IEEE Transactions on Neural Networks and Learning Systems}, vol. PP, pp. 1-17, 07/02 2018, doi: 10.1109/TNNLS.2018.2844093.
\bibitem{predictive_exit}
Li, Xiangjie, et al. "Predictive exit: Prediction of fine-grained early exits for computation-and energy-efficient inference." Proceedings of the AAAI Conference on Artificial Intelligence. Vol. 37. No. 7. 2023.

\bibitem{park2015} E. Park et al., "Big/little deep neural network for ultra low power inference," in \textit{Proceedings of the 10th International Conference on Hardware/Software Codesign and System Synthesis}, 2015: IEEE Press, pp. 124-132.

\bibitem{vitis_ai}
Xilinx Inc., \textit{Vitis AI: Development Stack for AI Inference on Xilinx Devices}. Available: \url{https://www.xilinx.com/products/design-tools/vitis/vitis-ai.html}

\bibitem{hls4ml}
J.~Duarte, S.~Summers, E.~Kreinar, et al., ``hls4ml: An Open-Source Codesign Workflow to Empower Scientific Low-Power Machine Learning Devices,'' \textit{Frontiers in Big Data}, vol.~3, 2021. DOI: \href{https://doi.org/10.3389/fdata.2020.600664}{10.3389/fdata.2020.600664}

\bibitem{tvm}
T.~Chen, T.~Moreau, Z.~Jiang, et al., ``TVM: An Automated End-to-End Optimizing Compiler for Deep Learning,'' in \textit{13th USENIX Symposium on Operating Systems Design and Implementation (OSDI '18)}, 2018. Available: \url{https://tvm.apache.org}

\bibitem{openvino}
Intel Corporation, ``OpenVINO™ Toolkit: Optimize, tune, and run AI inference,'' \url{https://www.intel.com/content/www/us/en/developer/tools/openvino-toolkit/overview.html}, 2023.




\bibitem{mazouz2020} A. Mazouz and C. P. Bridges, "Automated Offline Design-Space Exploration and Online Design Reconfiguration for CNNs," in \textit{2020 IEEE Conference on Evolving and Adaptive Intelligent Systems (EAIS)}, 27-29 May 2020 2020, pp. 1-9, doi: 10.1109/EAIS48028.2020.9122697.

\bibitem{mazouz2019ahs} A. Mazouz and C. P. Bridges, "Adaptive Hardware Reconfiguration for Performance Tradeoffs in CNNs," in \textit{2019 NASA/ESA Conference on Adaptive Hardware and Systems (AHS)}, 22-24 July 2019 2019, pp. 33-40, doi: 10.1109/AHS.2019.000-3.

\bibitem{siracusa2021} M. Siracusa et al., "A Comprehensive Methodology to Optimize FPGA Designs via the Roofline Model," \textit{IEEE Transactions on Computers}, pp. 1-1, 2021, doi: 10.1109/TC.2021.3111761.

\bibitem{cardoso2017} J. M. P. Cardoso, J. G. F. Coutinho, and P. C. Diniz, "Chapter 8 - Additional topics," in \textit{Embedded Computing for High Performance}, J. M. P. Cardoso, J. G. F. Coutinho, and P. C. Diniz Eds. Boston: Morgan Kaufmann, 2017, pp. 255-280.

\bibitem{konak2006} A. Konak, D. W. Coit, and A. E. Smith, "Multi-objective optimization using genetic algorithms: A tutorial," \textit{Reliability Engineering \& System Safety}, vol. 91, no. 9, pp. 992-1007, 2006/09/01/ 2006, doi: https://doi.org/10.1016/j.ress.2005.11.018.

\bibitem{lecun1998} Y. LeCun, L. Bottou, Y. Bengio, and P. Haffner, "Gradient-based learning applied to document recognition," \textit{Proceedings of the IEEE}, vol. 86, no. 11, pp. 2278-2324, 1998.

\bibitem{krizhevsky2012} A. Krizhevsky, "Learning Multiple Layers of Features from Tiny Images," University of Toronto, 05/08 2012.

\bibitem{netzer2011} Y. Netzer, T. Wang, A. Coates, A. Bissacco, B. Wu, and A. Ng, "Reading Digits in Natural Images with Unsupervised Feature Learning," NIPS, 01/01 2011.


\bibitem{mlperf} MLCommons, ``MLPerf Benchmark Suite,'' 2023. [Online]. Available: \url{https://mlcommons.org/en/}

\bibitem{raspi4} Raspberry Pi Foundation, ``Raspberry Pi 4 Model B,'' 2023. [Online]. Available: \url{https://www.raspberrypi.com/products/raspberry-pi-4-model-b/}

\bibitem{movidius1} Intel Corporation, ``Intel Movidius Neural Compute Stick,'' 2023. [Online]. Available: \url{https://www.intel.com/content/www/us/en/products/sku/97965/intel-movidius-neural-compute-stick/specifications.html}

\bibitem{movidius2} Intel Corporation, ``Intel Movidius Neural Compute Stick 2,'' 2023. [Online]. Available: \url{https://software.intel.com/en-us/openvino-toolkit}

\bibitem{nano} NVIDIA Corporation, ``Jetson Nano Developer Kit,'' 2023. [Online]. Available: \url{https://developer.nvidia.com/embedded/jetson-nano-developer-kit}

\bibitem{tx2} NVIDIA Corporation, ``Jetson TX2 Module,'' 2023. [Online]. Available: \url{https://developer.nvidia.com/embedded/jetson-tx2}

\bibitem{xaviernx} NVIDIA Corporation, ``Jetson Xavier NX Module,'' 2023. [Online]. Available: \url{https://developer.nvidia.com/embedded/jetson-xavier-nx}

\bibitem{xavier} NVIDIA Corporation, ``Jetson AGX Xavier Module,'' 2023. [Online]. Available: \url{https://developer.nvidia.com/embedded/jetson-agx-xavier}

\bibitem{tinker} ASUS IoT, ``Tinker Edge R,'' 2023. [Online]. Available: \url{https://www.asus.com/us/IoT/Tinker-Edge-R/}

\bibitem{coral} Google, ``Coral Dev Board,'' 2023. [Online]. Available: \url{https://coral.ai/products/dev-board/}

\bibitem{snapdragon} Qualcomm Technologies, ``Snapdragon 888 Mobile Platform,'' 2023. [Online]. Available: \url{https://www.qualcomm.com/products/snapdragon-888-5g-mobile-platform}

\bibitem{forgetting}
van de Ven, Gido M., Nicholas Soures, and Dhireesha Kudithipudi. "Continual learning and catastrophic forgetting." arXiv preprint arXiv:2403.05175 (2024).



\bibitem{clk_gating}
Guo, Meng, et al. "A RISC-V Domain-Specific Processor for Deep Learning-Based Channel Estimation." IEEE Transactions on Circuits and Systems I: Regular Papers (2025).

\bibitem{decaluwe2004} J. Decaluwe, "MyHDL: a python-based hardware description language," \textit{Linux J.}, vol. 2004, no. 127, p. 5, 2004.

\bibitem{haglund2003} P. Haglund, O. Mencer, W. Luk, and B. Tai, "PyHDL: Hardware Scripting with Python," in \textit{Engineering of Reconfigurable Systems and Algorithms}, 2003.

\end{thebibliography}
\bibliographystyle{IEEEtran}  

\vspace{-15mm} 
\begin{IEEEbiography}[{\includegraphics[width=1in,height=1.25in,clip,keepaspectratio]{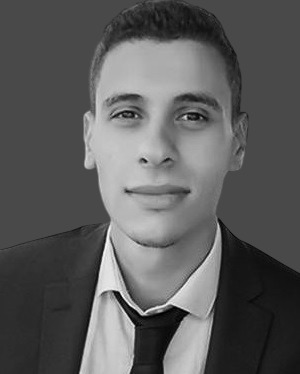}}]{Alaa Mazouz}
Alaa Mazouz received a B.S. from the institute of Electrical and Electronic Engineering at the university of Boumerdes, Algeria in 2014. An MSc in Power Engineering from the same institute in 2016. Recieved a PhD from  the Surrey Space Centre (SSC) University of Surrey in 2021, currently a Postdoc at Télécom Paris, COMELEC department working on embedded learned image compression, continual learning and runtime adaptive deep learning for embedded systems.
\end{IEEEbiography}

\vspace{-16mm} 
\begin{IEEEbiography}[{\includegraphics[width=1in,height=1.25in,clip,keepaspectratio]{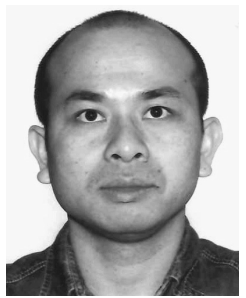}}]{Duc Han LE}
received his B.S. in Electronics and Telecom from Le Quy Don Technical Uni., Vietnam in 2005, and his M.Sc. in Communication Engineering from RWTH Aachen University, Germany in 2012. He earned his Ph.D. in Electronics and Telecom from Télécom Paris, France in 2015.
With over a decade of experience spanning academia and industry, he has held key R\&D roles at Synopsys, APEX Technologies, Télérad S.A.S \& PolyTech Nantes, NXP semiconductor \& Télécom Paris. He is currently a Research Engineer at Télécom Paris, France.
His research interests include embedded AI, continual learning, and EEG foundation models.
\end{IEEEbiography}

\vspace{-15mm} 
\begin{IEEEbiography}[{\includegraphics[width=1in,height=1.25in,clip,keepaspectratio]{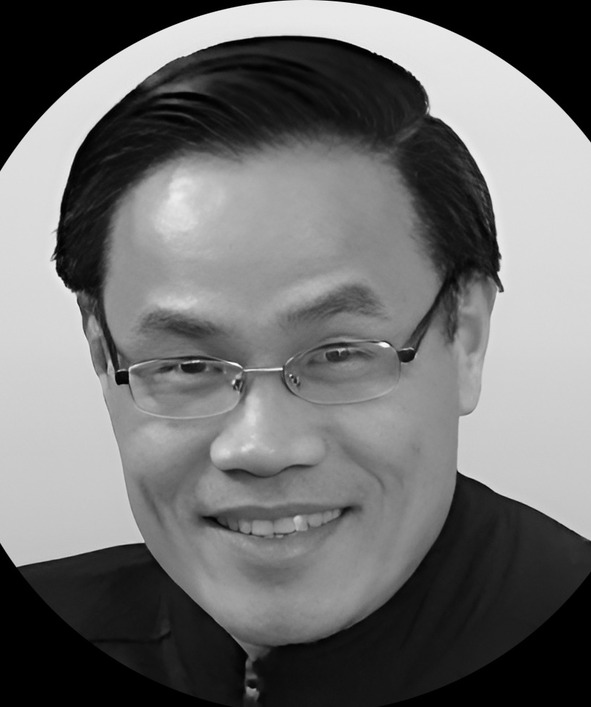}}]{Van Tam Nguyen} received the Diplôme d’Ingénieur from CentraleSupélec and an M.Sc. degree from Université Paris-Saclay, as well as a graduation degree in Image Processing from EPFL, Switzerland, in 2000. He earned his Ph.D. from Télécom Paris in 2004 and his Habilitation (H.D.R.) from Sorbonne University in 2016.
 
Since 2005, he has been with Télécom Paris, where he is currently a Full Professor, Head of the Computer Science and Networks Department, and a member of the Board of Directors. He also serves as Director of the ICMS Chair (Intelligent Cybersecurity for Mobility Systems) and Deputy Director of the LTCI laboratory.
 
At the Institut Polytechnique de Paris, he is the Acting Chair of the Computer Science, Data, and AI Department, as well as a member of the Teaching and Research Committee.
Internationally, he was a Rank A Guest Researcher at Japan’s NICT (2012–2013), working on game theory for security in cognitive radio networks. He held visiting appointments at UC Berkeley (2013–2016), where he introduced the CogniCom paradigm - a brain-inspired software-hardware approach to optimize IoT growth by bringing computing closer to the user through Smart Application Gateways and cloud computing - and at Stanford University (2016–2017), where he led the "Learning for Robust Communications" project.
Beyond academia, he served as Director of the Intek Institute (2018) and CEO of TAM-AI Co. Ltd. (2019–2021).
His research interests include Continual Learning, Adaptive AI for Networks, AI for Cybersecurity, and AI for Cognitive Science and Neuroscience. He has authored or co-authored over 100 publications, holds five patents, and has contributed to one major technology transfer. In recognition of his work, he received a Senior Marie Curie Fellowship from the European Commission in 2015.
\end{IEEEbiography}

\end{document}